\documentclass[namedreferences,hyperref,optionalrh]{spr-sola}
\usepackage{graphicx}        
\usepackage{amssymb}        
\usepackage{color}           
\usepackage{pdflscape}
\usepackage{caption}

\newcommand{\aap}{{\it Astron. Astrophys.}}

\newcommand{\apj}{{\it Astrophys. J.}}

\newcommand{\jgr}{{\it J. Geophys. Res.}}
\newcommand{\mnras}{{\it Mon. Not. R. Astron. Soc.}}

\newcommand{\solphys}{{\it Sol. Phys.}}

\chardef\us=`\_

\begin{document}

\begin{frontmatter}
\title{Turbulent power: a discriminator between sheaths and CMEs}

\author[addressref={aff1,aff2},email={deep.ghuge@students.iiserpune.ac.in}]{\inits{}\fnm{Deep}~\snm{Ghuge}}
\author[addressref=aff3,email={debesh.bhattacharjee@glasgow.ac.uk}]{\inits{}\fnm{Debesh}~\snm{Bhattacharjee}\orcid{[0000-0002-2651-5120]}}

\author[addressref=aff1,corref,email={p.subramanian@iiserpune.ac.in}]{\inits{}\fnm{Prasad}~\snm{Subramanian}}
\address[id=aff1]{Indian Institute of Science Education and Research, Dr Homi Bhabha Road, Pashan, Pune - 411008, India}
\address[id=aff2]{Catholic University, Washington, DC, USA}
\address[id=aff3]{School of Physics and Astronomy, University of Glasgow, Glasgow, G12 8QQ, UK}

\runningauthor{Ghuge, Bhattacharjee and Subramanian}
\runningtitle{Turbulent power in sheaths and CMEs}

\begin{abstract}
Solar coronal mass ejections (CMEs) directed at the Earth often drive large geomagnetic storms. 
Here we use velocity, magnetic field and proton density data from 152 CMEs that were sampled in-situ at 1 AU by the WIND spacecraft. We Fourier analyze fluctuations of these quantities in the quiescent pre-CME solar wind, sheath and magnetic cloud. We quantify the extent by which the power in turbulent (magnetic field, velocity and density) fluctuations in the sheath exceeds that in the solar wind background and in the magnetic cloud. For instance, the mean value of the power per unit volume in magnetic field fluctuations in the sheath is 76.7 times that in the solar wind background, while the mean value of the power per unit mass in velocity fluctuations in the sheath is 9 times that in the magnetic cloud. Our detailed results show that the turbulent fluctuation power is a useful discriminator between the ambient solar wind background, sheaths and magnetic clouds and can serve as a useful input for space weather prediction. 
\end{abstract}
\end{frontmatter}

\section{Introduction}
     \label{S-Introduction} 

The properties of Earth directed solar coronal mass ejections (CMEs) are a subject of considerable interest, owing to their dominant role in triggering space weather disturbances. Remote-sensing observations of CMEs are generally restricted to heliocentric distances of a few tens of $R_{\odot}$, and Earth-directed CMEs are especially difficult to observe with this technique. Near-Earth {\em in-situ} observations provide detailed measurements of CME properties, although they sample only the part of the CME which is along the line of intercept of the spacecraft. Such {\em in-situ} observations of CMEs have contributed to our understanding of their bulk properties (e.g., \citealp{1982Klein}, \citealp{2002hu}, \citealp{2009Mostl}, \citealp{2019NCSoPh}, \citealp{2002cid} and others) and to comprehensive interplanetary CME (ICME) catalogs (\citealp{2018NCSo}, \url{https://wind.nasa.gov/ICMEindex.php} and \citealp{richcane_2024}). 
Using {\em in-situ} observations, interplanetary CMEs are typically characterized by an increase in the magnetic field strength, a smooth rotation in the magnetic field direction and a decrease in proton temperature \citep{1982Klein}. CME sheaths have received considerable attention (e.g., \citealp{2017kilpua}, \citealp{2019kilpuasolar}, \citealp{2022Temmer}, \citealp{2024Larrodera}, \citealp{2020salman}), since it comprises the plasma near the leading edge of the CME and often determines the onset and severity of space weather disturbances. 

The characterizations of CMEs and their sheaths mentioned above typically refer to properties of the large-scale or average quantities; there is no appeal, for instance, to fluctuations in the magnetic field strength. 
We now turn our attention to fluctuations in the magnetic field, density and velocity. It is well recognized that the background solar wind comprises turbulent fluctuations in addition to the large-scale mean field (\citealp{2013Bruno}) and interiors of CMEs are also known to be turbulent (\citealp{2000Manoharan}, \citealp{2021Sorriso}, \citealp{2006Liu}, \citealp{2023Debesh}). CME sheaths are known to be characterized by enhanced magnetic field fluctuations (\citealp{2017kilpua}, \citealp{2019kilpuasolar}). However, the properties of velocity and density fluctuations in sheaths have not yet been studied, to the best of our knowledge.

  Turbulent fluctuations mediate the interaction between CMEs and the Earth's magnetosphere (\citealp{2003borovsky}). There are indications that turbulence can influence the rate of reconnection \citep{1999Lazarian}. This could mean that the turbulence levels in the sheath can influence the rate of reconnection between the sheath and the Earth's magnetic field, impacting the onset time and intensity of geomagnetic storms. The characteristics of turbulent fluctuations in CMEs and their sheaths are thus of interest from several perspectives.

Using moving box averages, \cite{2023Debesh} have studied the modulation indices of the magnetic field ($\delta B/B$), and density ($\delta n/n$) fluctuations. In the definition of the magnetic field modulation index, the quantity $B$ refers to the average value of the magnetic field magntitude in the moving box, while $\delta B$ refers to the root-mean-square (rms) value of the fluctuations in the magnetic field magnitude in the box. The modulation index for density fluctuations is defined similarly. They found that the values of $\delta B/B$ in CMEs were $\approx 0.79$ times that in the solar wind background. The corresponding ratio for $\delta n/n$ was 1.78. Here we take a different approach and study fluctuations using short-time Fourier transform (STFT) periodograms.

The rest of the paper is organized as follows: \S~\ref{S:DA} gives a short overview of STFTs and describes the dataset we use. \S~\ref{S:Results} describes our results concerning turbulent power and it's utility in distinguishing between the background solar wind, sheaths and CMEs. We summarize our results and draw conclusions in \S~\ref{S:summary}.

\section{Data Analysis} 
\label{S:DA}
\subsection{Short-time Fourier transforms (STFT)}
\label{S:STFT}
The STFT is a well known concept (e.g., \citealp{2001Gröchenig}). 
We mention the essential aspects here for completeness. While the usual Fourier transform of a function $a(t)$ is given by $\tilde{A}(f) = \int_{-\infty}^{\infty} a(t) \exp (i f t) dt$, the STFT is be given by
\begin{equation}
\hat{A}(f) = \int_{T_{\rm min}}^{T_{\rm max}} a(t) \exp (i f t) dt
\label{stftdef}
\end{equation}
The only difference between the usual Fourier transform and the STFT is as follows: the limits of integration for the usual Fourier transform are $(-\infty, \infty)$ while those for the STFT are finite $(T_{\rm min}, T_{\rm max})$; hence the adjective ``short time''. Equivalently, one can retain the same limits of integration as in the usual Fourier transform and multiply the integrand with a rectangular window function to get the STFT. The STFT periodogram of $a(t)$ refers to the quantity $|\hat{A}(f)|^{2}$, which is often called the power spectral density (e.g., \citealp{2000Bracewell}). This nomenclature can often be misleading, and it is best to remember that $|\hat{A}(f)|^{2}$ has units of $a^2$ per frequency. For instance, if $a$ represented velocity, it's units would be m/s (in SI units) and $| \hat{A}(f)|^{2}$ would have units of ${\rm m^{2}/s^{2}}$ per Hz.
We compute STFT periodograms for the plasma velocity, magnetic field and proton density (separately) in the ambient solar wind, sheath and magnetic cloud. The integration limits $(T_{\rm min}, T_{\rm max})$ (Eq~\ref{stftdef}) would correspond to the start and end times of the structure being considered; for instance, for the sheath STFT periodogram, $T_{\rm min}$ and $T_{\rm max}$ would correspond to the start and end times of the sheath. 

\subsection{Dataset used}
\label{S:Dataset}
We use WIND/MFI data for the same set of CMEs used in \cite{2023Debesh}. These are all the events observed between 1995 and 2015 that are listed as Fr or Fr+ in \cite{2018NCSo} and \url{https://wind.nasa.gov/ICMEindex.php}. In selecting events that are listed as Fr or Fr+, we restrict our study to events where the magnetic clouds (MCs) presumably conform best to the expectations of a flux rope structure. For completeness, we have listed the events in Table~\ref{S - Table A}. We use the nomenclature of \cite{2019NCSoPh}, where the location marked ``ICME start'' denotes the start of the sheath and the part between ``magnetic cloud (MC) start'' and ``MC end'' is taken to be the body of the CME. 
The part between the ICME start and the MC start is called the sheath and the part between the MC start and the MC end is called the MC. 
The background is a (manually chosen) one day stretch of quiet solar wind in the 5 days preceding the event. 
The details of the criteria used to define the quiet background are mentioned in \S~2 of \cite{2023Debesh}. 
For completeness, we quote the criteria for choosing the one hour background stretch from \cite{2023Debesh}: ``The background is a 24-hour window in the 5 days preceding the event and satisfying the following conditions: a) the rms fluctuations of the solar wind velocity for this 24-hour period should not exceed 10\% of the mean value b) the rms fluctuations of the total magnetic field for this 24-hour period should not exceed 20\% of the mean value c) there are no magnetic field rotations. The first two criteria ensure that the chosen background is quiet. Criterion c) distinguishes the background from the MCs, because MCs are characterized by large rotations of magnetic field components and low plasma beta.''

We have avoided events whose pre-event backgrounds overlap. Our work concerns fluctuations in the magnetic field ($B$), velocity ($V$) and proton density ($n$). The quantity $B$ refers to the magnitude of the magnetic field in the WIND/MFI dataset; i.e., $B \equiv | {\mathbf B} |$. Similarly, $V$ refers to the magnitude of the solar wind velocity; i.e., $V \equiv | {\mathbf V} |$. We compute the short-time Fourier transform (STFT) of these quantities for the background, sheath and MC separately. Thus $\hat{B}_{\rm bg}$ denotes the STFT of the time series for $B$ for the entire (one day) background stretch, $\hat{B}_{\rm sheath}$ denotes the STFT of the time series for $B$ for the entire sheath region and $\hat{B}_{\rm MC}$ denotes the STFT of the time series for $B$ for the entire MC. The STFTs for the velocity ($\hat{V}_{\rm bg}$, $\hat{V}_{\rm sheath}$, $\hat{V}_{\rm MC}$) and density ($\hat{n}_{\rm bg}$, $\hat{n}_{\rm sheath}$, $\hat{n}_{\rm MC}$) are defined similarly. In computing $\hat{B}_{\rm bg}$, $\hat{V}_{\rm bg}$ and $\hat{n}_{\rm bg}$, $T_{\rm min}$ and $T_{\rm max}$ (Eq~\ref{stftdef}) correspond to the start and end times of the stretch of solar wind that we define as the background. Similarly, $T_{\rm min}$ and $T_{\rm max}$ would correspond to the start and end times of the sheath in computing $\hat{B}_{\rm sheath}$, $\hat{V}_{\rm sheath}$ and $\hat{n}_{\rm sheath}$, while they would correspond to the start and end times of the MC in computing $\hat{B}_{\rm MC}$, $\hat{V}_{\rm MC}$ and $\hat{n}_{\rm MC}$.
Our analysis makes use of the periodograms of $\hat{B}_{\rm bg}$, $\hat{B}_{\rm sheath}$, $\hat{B}_{\rm MC}$, $\hat{V}_{\rm bg}$, $\hat{V}_{\rm sheath}$, $\hat{V}_{\rm MC}$, $\hat{n}_{\rm bg}$, $\hat{n}_{\rm sheath}$ and $\hat{n}_{\rm MC}$.
\section{Results}
\label{S:Results}
\subsection{Durations of sheath and MC}
\label{S:dur}
\begin{figure}    
\centerline{\includegraphics[width=0.7\textwidth,clip=]{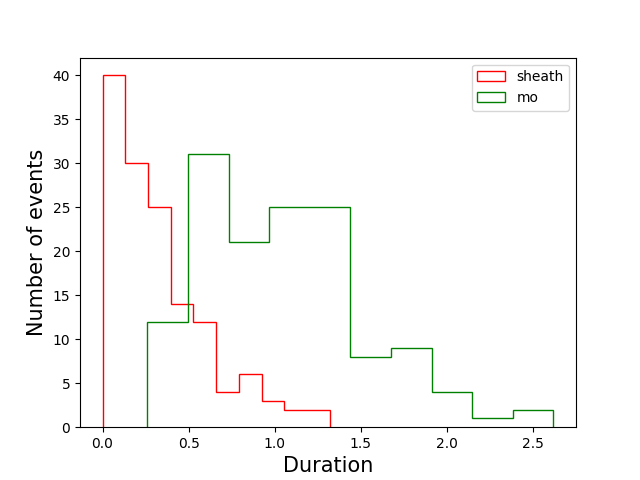}}
\small
        \caption{A histogram of the sheath and magnetic cloud durations (in days). The background duration is one day by definition. The mean sheath duration is 0.33 days and the median is 0.25 days. The mean MC duration is 1.04 days and the median is 0.99 days.}
\label{F-durations}
\end{figure}

Before describing our main results, we describe our findings regarding the durations of the background, sheath and MCs. The mean, median and standard deviation of the durations are shown in table~\ref{Durationstable}. A histogram of the durations is shown in Fig~\ref{F-durations}. As mentioned in \S~\ref{S:Dataset}, the duration of the solar wind background is chosen to be one day. 
These results are broadly consistent with the findings of \cite{2014Mitsakou}. 
\begin{table}
\begin{tabular}{ c  c  c  c }
       & \begin{tabular}[c]
       {@{}l@{}}Mean\\ (days)\end{tabular} & \begin{tabular}[c]{@{}l@{}}Median\\ (days)\end{tabular} & \begin{tabular}[c]{@{}l@{}}Standard Deviation\\ (days)\end{tabular} \\
       \hline
Sheath & 0.33                                                  & 0.255                                                   & 0.29                                                                \\
\hline 
MC     & 1.04                                                  & 0.99                                                    & 0.47                                                          \\
\hline
\end{tabular}
\captionof{table}{Durations of sheath and MC}
\label{Durationstable}
\end{table}

\subsection{Power in turbulent fluctuations: background, sheath and MC}

We define the area under the periodograms $E_{B\, {\rm bg}}$, $E_{B\, {\rm sheath}}$, $E_{B\, {\rm MC}}$, $E_{V\, {\rm bg}}$, $E_{V\, {\rm sheath}}$, $E_{V\, {\rm MC}}$, $E_{n\, {\rm bg}}$, $E_{n\, {\rm sheath}}$ and $E_{n\, {\rm MC}}$ as follows:

{\begin{eqnarray}
\nonumber
E_{B\, {\rm bg}} = \int_{f_{1}}^{f_{2}} | \hat{B}_{\rm bg} (f) |^{2}\, df \, , \,\, E_{B\, {\rm sheath}} = \int_{f_{1}}^{f_{2}} | \hat{B}_{\rm sheath} (f) |^{2}\,df \, , \,\, E_{B\, {\rm MC}} (f) = \int_{f_{1}}^{f_{2}} | \hat{B}_{\rm MC} (f) |^{2}\,df \, , \\
\nonumber
E_{V\, {\rm bg}} = \int_{f_{1}}^{f_{2}} | \hat{V}_{\rm bg} (f) |^{2}\, df \, , \,\, E_{V\, {\rm sheath}} = \int_{f_{1}}^{f_{2}} | \hat{V}_{\rm sheath} (f) |^{2}\,df \, , \,\, E_{V\, {\rm MC}} (f) = \int_{f_{1}}^{f_{2}} | \hat{V}_{\rm MC} (f) |^{2}\,df \, , \\
E_{n\, {\rm bg}} = \int_{f_{1}}^{f_{2}} | \hat{n}_{\rm bg} (f) |^{2}\, df \, , \,\, E_{n\, {\rm sheath}} = \int_{f_{1}}^{f_{2}} | \hat{n}_{\rm sheath} (f) |^{2}\,df \, , \,\, E_{n\, {\rm MC}} (f) = \int_{f_{1}}^{f_{2}} | \hat{n}_{\rm MC} (f) |^{2}\,df \,\,\,\,\,\,\,\,\, .
\label{eqE}
\end{eqnarray}}
Eq~\ref{eqE} involves integrals over frequency with lower and upper bounds on frequency (denoted by $f_{1}$ and $f_{2}$ respectively), which represent the lowest and highest frequencies of fluctuations in the respective Fourier spectrum (e.g., the spectrum of $\hat{B}_{\rm bg} (f)$ or that of $\hat{n}_{\rm sheath} (f)$). Although we use the same symbols ($f_{1}$ and $f_{2}$) for simplicity of notation, $f_{1}$ and $f_{2}$ are not the same numbers for all the spectra. $f_{1}$ is typically $3 \times 10^{-5}$ Hz and $f_{2}$ is typically $5 \times 10^{-3}$ Hz. By comparison, magnetic field and velocity spectra in the ambient solar wind typically extend from a few times $10^{-4}$ to a few times $10^{-3}$ Hz \citep{2012borovsky}.  Going by the units used for WIND/MFI data, the quantities $E_{B\, {\rm bg}}$, $E_{B\, {\rm sheath}}$ and $E_{B\, {\rm MC}}$ have units of ${\rm nT}^{2}$, and are proportional to energy per unit volume (${\rm J\,m^{-3}}$) in magnetic fluctuations. The quantities $E_{V\, {\rm bg}}$, $E_{V\, {\rm sheath}}$ and $E_{V\, {\rm MC}}$ have units of ${\rm km^{2}\,s^{-2}}$, and are proportional to kinetic energy per unit mass (${\rm J\,kg^{-1}}$) in velocity fluctuations. The quantities $E_{n\, {\rm bg}}$, $E_{n\, {\rm sheath}}$ and $E_{n\, {\rm MC}}$ have units of ${\rm cm^{-6}}$.

Characterizations of turbulent intensity typically involve measurements of the rms deviation of the magnetic field about the mean value inside a moving box of a fixed duration like one hour; e.g., \cite{2017kilpua}, \cite{2019kilpuasolar}. Similarly, \cite{2023Debesh} compute the ratio of the rms deviation of density and velocity to the mean inside a one hour moving box. The solar wind velocity and magnetic field spectra shown in \cite{2012borovsky} are computed from 4.55 hour intervals. \footnote{For the solar wind, several such spectra (from each 4.55 hour interval) are typically stacked together to get an average spectrum that has lower noise.} For our purposes, the point to note is that measures of turbulent fluctuations (whether they are rms deviations or fluctuation spectra) are evaluated on a per-unit-time interval basis. In our case, the appropriate measure would be the turbulent energy densities (defined in Eq~\ref{eqE}) per unit time. Accordingly, we define

{\begin{eqnarray}
\nonumber
P_{B\, {\rm bg}} = \frac{E_{B\, {\rm bg}}}{T_{\rm bg}}\, , \,\, P_{B\, {\rm sheath}} = \frac{E_{B\, {\rm sheath}}}{T_{\rm sheath}} \, , \,\, P_{B\, {\rm MC}} = \frac{E_{B\, {\rm MC}}}{T_{\rm MC}} \, , \\
\nonumber
P_{V\, {\rm bg}} = \frac{E_{V\, {\rm bg}}}{T_{\rm bg}}\, , \,\, P_{V\, {\rm sheath}} = \frac{E_{V\, {\rm sheath}}}{T_{\rm sheath}} \, , \,\, P_{V\, {\rm MC}} = \frac{E_{V\, {\rm MC}}}{T_{\rm MC}} \, , \\
P_{n\, {\rm bg}} = \frac{E_{n\, {\rm bg}}}{T_{\rm bg}}\, , \,\, P_{n\, {\rm sheath}} = \frac{E_{n\, {\rm sheath}}}{T_{\rm sheath}} \, , \,\, P_{n\, {\rm MC}} = \frac{E_{n\, {\rm MC}}}{T_{\rm MC}}
\label{eqP}
\end{eqnarray}}
where $T_{\rm bg}$, $T_{\rm sheath}$ and $T_{\rm MC}$ are the time durations of the background, sheath and MC respectively. These also provide a natural transition to comparisons with wavelet spectra. The quantities $P_{B\, {\rm bg}}$, $P_{B\, {\rm sheath}}$ and $P_{B\, {\rm MC}}$ are in units of ${\rm nT^{2}\,day^{-1}}$, which is proportional to the power per unit volume (${\rm W\,m^{-3}}$) in turbulent magnetic fluctuations in the background, sheath and MC respectively. The quantities $P_{V\, {\rm bg}}$, $P_{V\, {\rm sheath}}$ and $P_{V\, {\rm MC}}$ are in units of ${\rm km^{2}\,s^{-2}\,day^{-1}}$, which is proportional to the power per unit mass (${\rm W\,kg^{-1}}$) in turbulent velocity fluctuations in the background, sheath and MC respectively. 
While $P_{n\, {\rm bg}}$, $P_{n\, {\rm sheath}}$ and $P_{n\, {\rm MC}}$ are in units of ${\rm (number\,cm^{-3})^{2}}$ and cannot be identified with any such physical quantity, they are still useful diagnostics of density fluctuations. 

The quantities defined in Eq~\ref{eqP} provide clear discriminants between the sheath and the MC; their value(s) for each event are listed in Table~\ref{Table_comp}. The quantities in Eq~\ref{eqP} rely on the integrals in Eq~\ref{eqE}, which are computed via numerical integration after interpolating the time series of the integrands using the Mathematica software. The interpolation process often fails when the data is highly oscillatory 
and this has led us to reject 14 events from the original list in Table~\ref{S - Table A}, leading to only 138 events being listed in Table~\ref{Table_comp}. Of these, 21 events have the ICME start coinciding with the MC start, which means that $T_{\rm sheath} = 0$ for these events. Since $T_{\rm sheath}$ appears in the denominator in Eq~\ref{eqP}, we have not listed $P_{\rm B\,sheath}$, $P_{\rm V\,sheath}$ and $P_{\rm n\,sheath}$ for these 21 events in Table~\ref{Table_comp}. 

\begin{figure}    
\centerline{\includegraphics[width=0.7\textwidth,clip=]{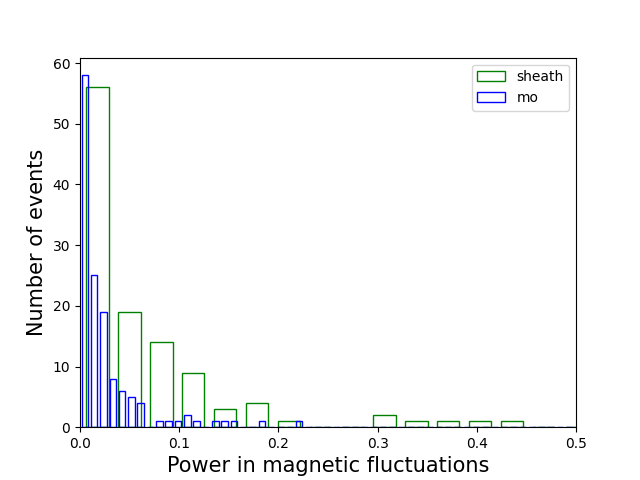}}
\small
        \caption{Histograms of the power per unit volume in magnetic field fluctuations for the sheath and MC (Eq~\ref{eqP}).}
\label{F-Bpower1}
\end{figure}

\begin{figure}    
\centerline{\includegraphics[width=0.7\textwidth,clip=]{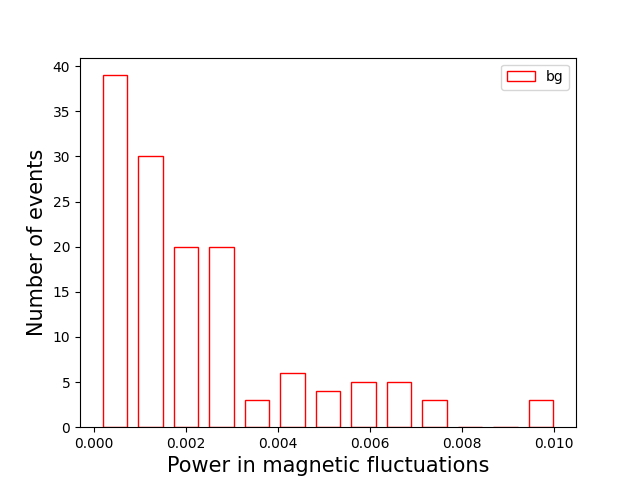}}
\small
        \caption{Histograms of the power per unit volume in magnetic field fluctuations for the solar wind background (Eq~\ref{eqP}).}
\label{F-Bpower2}
\end{figure}

\begin{figure}    
\centerline{\includegraphics[width=0.7\textwidth,clip=]{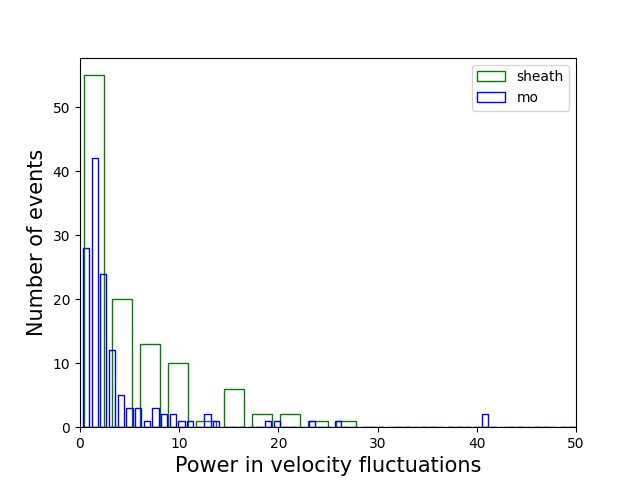}}
\small
        \caption{Histograms of the power per unit mass in plasma velocity fluctuations for the sheath and MC (Eq~\ref{eqP}).}
\label{F-Vpower1}
\end{figure}

\begin{figure}    
\centerline{\includegraphics[width=0.7\textwidth,clip=]{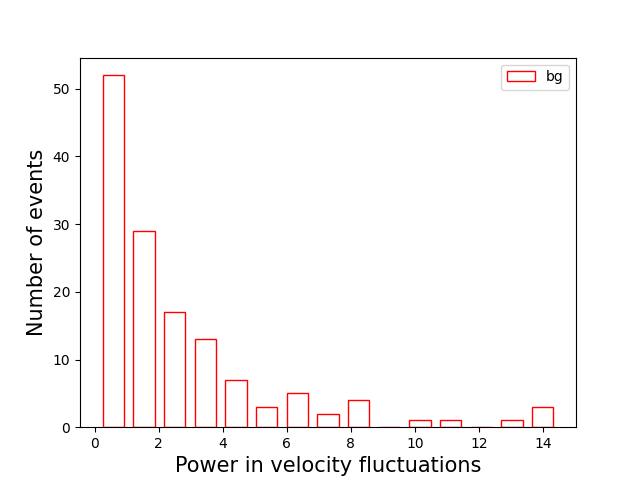}}
\small
        \caption{Histograms of the power per unit mass in plasma velocity fluctuations for the solar wind background.}
\label{F-Vpower2}
\end{figure}

\begin{figure}    
\centerline{\includegraphics[width=0.7\textwidth,clip=]{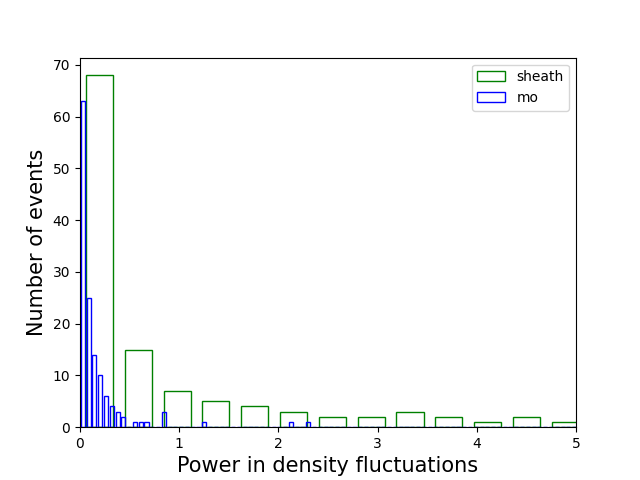}}
\small
        \caption{Histograms of the power density in density fluctuations for the sheath and MC (Eq~\ref{eqP}).}
\label{F-dpower1}
\end{figure}

\begin{figure}    
\centerline{\includegraphics[width=0.7\textwidth,clip=]{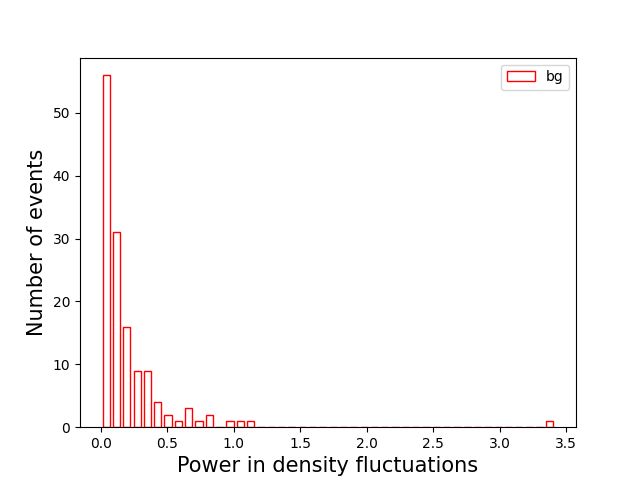}}
\small
        \caption{Histograms of the power density in density fluctuations for the solar wind background (Eq~\ref{eqP}).}
\label{F-dpower2}
\end{figure}

Histograms of the quantities defined in Eq~\ref{eqP} (using the events in table~\ref{Table_comp}) will provide a useful overview. Figure~\ref{F-Bpower1} is a histogram of  $P_{B\, {\rm sheath}}$ and $P_{B\, {\rm MC}}$, while figure~\ref{F-Bpower2} is a histogram of $P_{B\, {\rm bg}}$. Figure~\ref{F-Vpower1} is a histogram of $P_{V\, {\rm sheath}}$ and $P_{V\, {\rm MC}}$ while figure~\ref{F-Vpower2} is a histogram of $P_{V\, {\rm bg}}$. Figure~\ref{F-dpower1} is a histogram of $P_{n\, {\rm sheath}}$ and $P_{n\, {\rm MC}}$ while figure~\ref{F-dpower2} is a histogram of $P_{n\, {\rm bg}}$. The backgrounds are quiet (by choice); the magnetic field fluctuations in the background are typically two orders of magnitude smaller, velocity fluctuations smaller by a factor of two and a density fluctuations smaller by a factor of 0.7. We therefore show the histograms for the background(s) separately for clarity. 
We will use the mean, median and most probable values (MPVs) of these histograms for interpretation. Computing means and medians is a straightforward matter, but the MPV involves finding the peak of the envelope to the histogram, and is somewhat sensitive to the bin size. We have used the `auto' option in matplotlib to determine the optimum bin size for each histogram. This option chooses the smaller of the bin sizes recommended by the Sturges and the Freedman Diaconis estimators (see, for instance, \citealp{1997Wand01021997}). 

\begin{figure}    
\centerline{\includegraphics[width=0.7\textwidth,clip=]{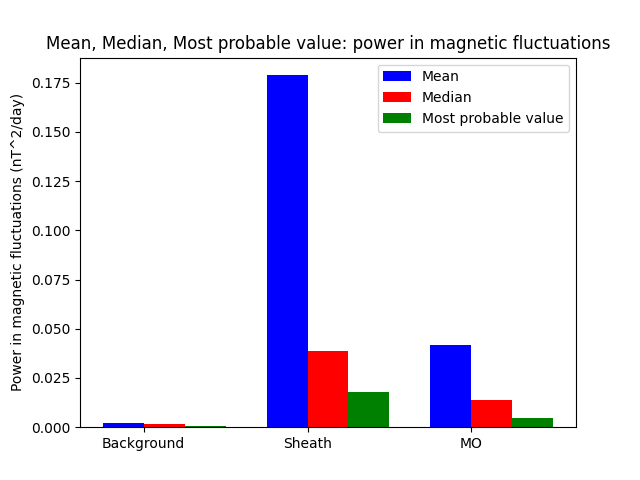}}
\small
        \caption{The mean, median and most probable value of the power per unit volume in magnetic fluctuations (Figs~\ref{F-Bpower1} and \ref{F-Bpower2}). All quantities are in ${\rm nT^{2}\,day^{-1}}$}
\label{F-barB}
\end{figure}

\begin{figure}    
\centerline{\includegraphics[width=0.7\textwidth,clip=]{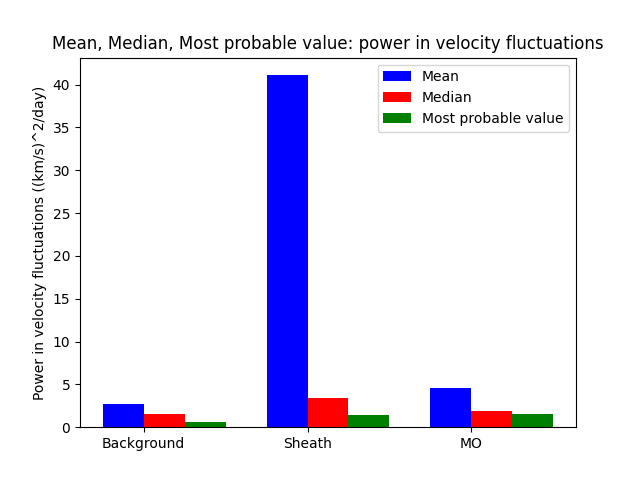}}
\small
        \caption{The mean, median and most probable value of the power per unit mass in velocity fluctuations (Figs~\ref{F-Vpower1} and \ref{F-Vpower2}). All quantities are in ${\rm km^{2}\,s^{-2}\,day^{-1}}$}
\label{F-barV}
\end{figure}

\begin{figure}    
\centerline{\includegraphics[width=0.7\textwidth,clip=]{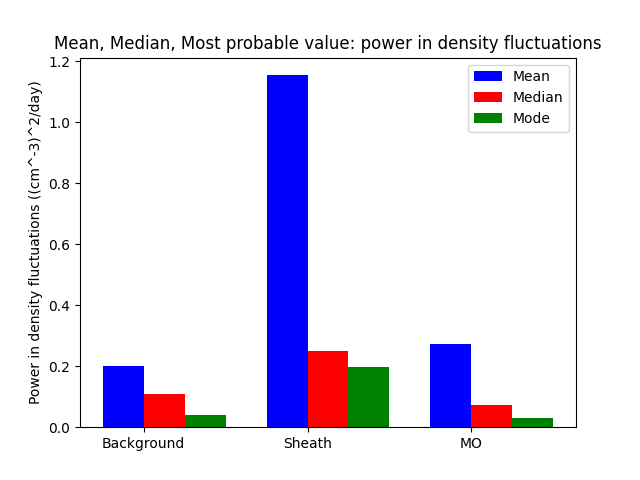}}
\small
        \caption{The mean, median and most probable value of the power density in proton density fluctuations (Figs~\ref{F-dpower1} and \ref{F-dpower2}). All quantities are in ${\rm cm^{-6}\,day^{-1}}$}
\label{F-barn}
\end{figure}

A comparison of the means, medians and most probable values (MPV) of figures \ref{F-Bpower1}, \ref{F-Bpower2}, \ref{F-Vpower1}, \ref{F-Vpower2}, \ref{F-dpower1} and \ref{F-dpower2} yields concrete information. Specifically, 
 \begin{itemize}
  \item Figure~\ref{F-barB} shows a comparison of the mean (blue bar), median (red bar) and MPV (green bar) of the histograms of figures~\ref{F-Bpower1} and \ref{F-Bpower2} (which show the power per unit volume (${\rm nT^{2}/day}$) in magnetic field fluctuations).

  \item 
  Figure~\ref{F-barV} shows a comparison of the mean (blue bar), median (red bar) and MPV (green bar) of the histograms of figure~\ref{F-Vpower1} and figure~\ref{F-Vpower2} (which show the power per unit mass (${\rm km^{2}\,s^{-2}\,per\,day}$) in plasma velocity fluctuations).
  \item 
Figure~\ref{F-barn} shows a comparison of the mean (blue bar), median (red bar) and MPV (green bar) of the histograms of figure~\ref{F-dpower1} and figure~\ref{F-dpower2}.  
\end{itemize} 
 
 Table~\ref{Table-combinedstats} shows the means, MPVs in the sheath in relation to those in the background and the MC. These values refer to the bar charts of figures~\ref{F-barB}, \ref{F-barV} and \ref{F-barn}. For instance, the quantity ${\rm Mean_{sheath}/Mean_{bg}}$ for $B$ would be the ratio of the blue bar for the sheath to the blue bar for the background in figure~\ref{F-barB}, the quantity ${\rm MPV_{sheath}/MPV_{MC}}$ for $n$ would be the ratio of the green bar for the sheath to the green bar for the MC in figure~\ref{F-barn} and so on. The values in rows 1, 3 and 5 of table~\ref{Table-combinedstats} are considerably larger than unity. In other words, (by way of the mean, median and the most probable value for $B$, $V$ and $n$) the sheath is significantly more turbulent than the background. The sheath is also more turbulent than the MC (since the values in rows 2, 4 and 6 of table~\ref{Table-combinedstats} are  $\gtrsim 1$), although the ratios are not as large, and the ${\rm MPV_{sheath}/MPV_{MC}}$ for $V$ is $\approx 1$. Previous studies such as \citet{2019kilpuasolar} have highlighted enhanced turbulence levels in the sheath, but they have only used magnetic field measurements. We have studied turbulent fluctuations in the magnetic field, velocity and density, which allow us to be more comprehensive in finding discriminators between the solar wind background, sheath and MC. Furthermore, they have used rms fluctuations about the mean in a moving box, which is a different technique from what we have used here.

 The numbers cited in table~\ref{Table-combinedstats} give an overview of all the criteria that could possibly be used to distinguish between the solar wind background, sheath and MC. We now discuss the most significant of these criteria. If one were to go only by the means cited in table~\ref{Table-combinedstats}, the value of 76.7 in row 1, column 2 suggests that magnetic field fluctuations are the best discriminator between the sheath and the background. Going by the same criterion, the value of 9 in row 2, column 3 of table~\ref{Table-combinedstats} suggests that velocity fluctuations are the best discriminator between the sheath and the MC. Given the rather skewed nature of the histograms (figures~\ref{F-Bpower1}, \ref{F-Bpower2}, \ref{F-Vpower1}, \ref{F-Vpower2}, \ref{F-dpower1} and \ref{F-dpower2}), it may be argued that the most probable values (MPVs) might be more relevant, since the means are influenced by the outliers. Going by the MPVs, the value of 39.2 in row 5, column 2 of table~\ref{Table-combinedstats} suggests that magnetic field fluctuations are the best discriminator between the sheath and the background. However, (going by MPVs) the value of 6.7 in row 6, column 4 of table~\ref{Table-combinedstats} suggests that density fluctuations are the best discriminator between the sheath and the MC.

\section{Summary and Discussion}
\label{S:summary}
\subsection{Summary of results}
We study the turbulent fluctuation characteristics of the solar wind backgrounds, sheath and CMEs using a database of 152 well observed events. Our dataset comprises near-Earth {\em in-situ} measurements of proton density ($n$) and (the magnitudes of) plasma velocity ($V$) and magnetic field ($B$). This provides a sampling of the plasma properties along the line of intercept of WIND spacecraft. In keeping with the convention followed in the WIND ICME database, the start of the event is called the ICME start and the magnetically structured part of the event that conforms to a flux rope structure is called the magnetic cloud (MC). The part of the event between the ICME start and the MC start is called the sheath. The background is a stretch of quiet solar wind prior to the start of the event. We find that sheaths last for $\approx 0.33$ days on average and MCs last for $\approx 1$ day (figure~\ref{F-durations}).

We compute separate STFT periodograms for the background, sheath and MC (respectively) and use it to find the power density in turbulent $B$, $V$ and $n$ fluctuations in each of these regions (Eq~\ref{eqP}). Histograms of these quantities are shown in figures \ref{F-Bpower1}, \ref{F-Bpower2}, \ref{F-Vpower1}, \ref{F-Vpower2}, \ref{F-dpower1} and \ref{F-dpower2} and their statistics are summarized in figures~\ref{F-barB}, \ref{F-barV} and \ref{F-barn}. The ratio of the heights of the bars corresponding to the sheath, solar wind background and MC in figures~\ref{F-barB}, \ref{F-barV} and \ref{F-barn} are given in table~\ref{Table-combinedstats}. 

\begin{table}
\begin{tabular}{cccc}
                                    & $B$  & $V$  & $n$ \\
                                    \hline
${\rm Mean_{sheath}/Mean_{bg}}$     &  76.7 & 15.4 & 5.7  \\
${\rm Mean_{sheath}/Mean_{MC}}$     & 4.3  & 9    & 4.2 \\
${\rm Median_{sheath}/Median_{bg}}$ & 23.8 & 2.1  & 2.3 \\
${\rm Median_{sheath}/Median_{MC}}$ & 2.8  & 1.8  & 3.5 \\
${\rm MPV_{sheath}/MPV_{bg}}$       & 39.2 & 2.4  & 5   \\
${\rm MPV_{sheath}/MPV_{MC}}$       & 3.6  & 0.94 & 6.7 \\
\hline
\captionof{table}{Contrast between turbulent power in the sheath, background and MC}\label{Table-combinedstats}
\end{tabular}
\end{table}

The most significant discriminators (by way of turbulent fluctuations) are as follows: the mean value of the power per unit volume in magnetic field fluctuations (${\rm nT^{2}\,day^{-1}}$) in the sheath is 76.7 times that in the solar wind background, while the mean value of the power per unit mass in velocity fluctuations (${\rm km^{2} \, s^{-2}\,day^{-1}}$) in the sheath is 9 times that in the MC. Going by the long tails of the histograms in figures \ref{F-Bpower1}, \ref{F-Bpower2}, \ref{F-Vpower1}, \ref{F-Vpower2}, \ref{F-dpower1} and \ref{F-dpower2}, it may be argued that the means are influenced by the infrequent, large values, and most probable values (MPVs) are better statistical measures. The most probable value of the magnetic fluctuation power per unit volume (${\rm nT^{2}\,day^{-1}}$) in the sheath is 39.2 times that in the solar wind background. 
Interestingly, by way of the MPVs, the density fluctuation power per unit volume in the sheath is 6.7 times that in the MC, and is thus the best discriminator between these regions. 
\subsection{Discussion and conclusions}
We show a simple, practical method to quantify the notion of i) the sheath being significantly more turbulent than the background solar wind and ii) the sheath being a turbulent precursor to a relatively quieter MC. 
It is already well known that magnetic field fluctuations in the sheath are larger than those in the ambient solar wind \citep{2017kilpua}. 

We build on this and give comprehensive quantitative measures of the contrast between the turbulent power in the sheath, solar wind background and MC, not only for the magnetic field, but also for the solar wind speed and the proton density.
The data in table~\ref{Table-combinedstats} can serve as useful inputs to automated methods (which often have trouble distinguishing between the sheath and MC; e.g., \citealp{2024Narock}) for determining the start of the sheath and the transition between the sheath and the MC. 
 For instance, if the mean turbulent power in magnetic fields increases by a factor of $\approx 76$, that in the velocity by a factor of $\approx 15$ and that in the density by a factor of $\approx 5$, it indicates that the spacecraft has transitioned from the background solar wind to the sheath region. Similarly, if the mean turbulent power in magnetic fields decreases by a factor of $\approx 1/4.3 = 0.23$, that in the velocity by a factor of $\approx 1/9 = 0.1$ and that in the density by a factor of $\approx 1/4.2 = 0.23$, it indicates that the spacecraft has transitioned from the sheath into the MC. The numbers in table~\ref{Table-combinedstats} are probably lower bounds, because we consider each region (e.g., sheath, MC) to be a single entity and our results only give average values for the entire sheath/MC. The turbulent fields in these regions likely have substructure, with power concentrated in turbulent ``spots''. The substructure  can be revealed via techniques such as wavelet transforms \citep{2013kilpua} and our results can be viewed as a precursor to more detailed wavelet-based analysis. 
Furthermore, we have used only with the magnitudes of the plasma velocity and magnetic field. A separate examination of their components might also be useful. In particular, the role of turbulent fluctuations in the southward directed magnetic field component in causing geomagnetic disturbances might be interesting.

\begin{acks}
 The authors thank the anonymous referee for insightful comments and suggestions that have helped improve this paper.
\end{acks}


\appendix   
\renewcommand{\thetable}{A\arabic{table}}
\setcounter{table}{0}
\begin{table}
%
  
\captionof{table}{
The list of the 152 Wind ICME events we use in this study. The arrival date and time of the ICME at the position of Wind measurement and the arrival and departure dates \& times of the associated magnetic clouds (MCs) are taken from Wind ICME catalogue (\url{https://wind.nasa.gov/ICMEindex.php}). Fr events indicate MCs with a single magnetic field rotation between $90^{\circ}$ and $180^{\circ}$ and F+ events indicate MCs with a single magnetic field rotation greater than $180^{\circ}$ \citep{2018NCSo}.} The 14 events marked with and asterisk (*) coincide with the near earth counterparts of 14 CMEs listed in \citep{2017Nishtha}. \label{S - Table A}

  \begin{center}
  
  \begin{tabular}{cccclccc}
  
   \hline\hline
   CME        & CME Arrival date & MC start  & MC end       & Flux rope \\
   event      & and time[UT]   &  date and   & date and     & type   \\
    number    &  (1AU)         &  time [UT]   &   time [UT]   &      \\
    \hline
 
 1    &  1995 03 04 , 00:36 & 1995 03 04 , 11:23 & 1995 03 05 , 03:06 & Fr \\
 2    &  1995 04 03 , 06:43 & 1995 04 03 , 12:45 & 1995 04 04 , 13:25 & F+ \\
 3    &  1995 06 30 , 09:21 & 1995 06 30 , 14:23 & 1995 07 02 , 16:47 & Fr \\
 4    &  1995 08 22 , 12:56 & 1995 08 22 , 22:19 & 1995 08 23 , 18:43 & Fr \\
 5    &  1995 09 26 , 15:57 & 1995 09 27 , 03:36 & 1995 09 27 , 21:21 & Fr \\
 6    &  1995 10 18 , 10:40 & 1995 10 18 , 19:11 & 1995 10 20 , 02:23 & Fr \\
 7    &  1996 02 15 , 15:07 & 1996 02 15 , 15:07 & 1996 02 16 , 08:59 & F+ \\
 8    &  1996 04 04 , 11:59 & 1996 04 04 , 11:59 & 1996 04 04 , 21:36 & Fr \\ 
 9    &  1996 05 16 , 22:47 & 1996 05 17 , 01:36 & 1996 05 17 , 11:58 & F+ \\
 10   &  1996 05 27 , 14:45 & 1996 05 27 , 14:45 & 1996 05 29 , 02:22 & Fr \\
 11   &  1996 07 01 , 13:05 & 1996 07 01 , 17:16 & 1996 07 02 , 10:17 & Fr \\
 12   &  1996 08 07 , 08:23 & 1996 08 07 , 11:59 & 1996 08 08 , 13:12 & Fr \\
 13   &  1996 12 24 , 01:26 & 1996 12 24 , 03:07 & 1996 12 25 , 11:44 & F+ \\ 
 14   &  1997 01 10 , 00:52 & 1997 01 10 , 04:47 & 1997 01 11 , 03:36 & F+ \\
 15   &  1997 04 10 , 17:02 & 1997 04 11 , 05:45 & 1997 04 11 , 19:10 & Fr \\
 16   &  1997 04 21 , 10:11 & 1997 04 21 , 11:59 & 1997 04 23 , 07:11 & F+ \\
 17   &  1997 05 15 , 01:15 & 1997 05 15 , 10:00 & 1997 05 16 , 02:37 & F+ \\
 18   &  1997 05 26 , 09:09 & 1997 05 26 , 15:35 & 1997 05 28 , 00:00 & Fr \\
 19   &  1997 06 08 , 15:43 & 1997 06 09 , 06:18 & 1997 06 09 , 23:01 & Fr \\
 20   &  1997 06 19 , 00:00 & 1997 06 19 , 05:31 & 1997 06 20 , 22:29 & Fr \\
 21   &  1997 07 15 , 03:10 & 1997 07 15 , 06:48 & 1997 07 16 , 11:16 & F+ \\
 22   &  1997 08 03 , 10:10 & 1997 08 03 , 13:55 & 1997 08 04 , 02:23 & Fr \\
 23   &  1997 08 17 , 01:56 & 1997 08 17 , 06:33 & 1997 08 17 , 20:09 & Fr \\
 24   &  1997 09 02 , 22:40 & 1997 09 03 , 08:38 & 1997 09 03 , 20:59 & Fr \\
 25   &  1997 09 18 , 00:30 & 1997 09 18 , 04:07 & 1997 09 19 , 23:59 & F+ \\
 26   &  1997 10 01 , 11:45 & 1997 10 01 , 17:08 & 1997 10 02 , 23:15 & Fr \\
 27   &  1997 10 10 , 03:08 & 1997 10 10 , 15:33 & 1997 10 11 , 22:00 & F+ \\
 28   &  1997 11 06 , 22:25 & 1997 11 07 , 06:00 & 1997 11 08 , 22:46 & F+ \\
 29   &  1997 11 22 , 09:12 & 1997 11 22 , 17:31 & 1997 11 23 , 18:43 & F+ \\
 30   &  1997 12 30 , 01:13 & 1997 12 30 , 09:35 & 1997 12 31 , 08:51 & Fr \\
 31   &  1998 01 06 , 13:29 & 1998 01 07 , 02:23 & 1998 01 08 , 07:54 & F+ \\
 32   &  1998 01 28 , 16:04 & 1998 01 29 , 13:12 & 1998 01 31 , 00:00 & F+ \\
 33   &  1998 03 25 , 10:48 & 1998 03 25 , 14:23 & 1998 03 26 , 08:57 & Fr \\
 34   &  1998 03 31 , 07:11 & 1998 03 31 , 11:59 & 1998 04 01 , 16:18 & Fr \\
 35   &  1998 05 01 , 21:21 & 1998 05 02 , 11:31 & 1998 05 03 , 16:47 & Fr \\
 36   &  1998 06 02 , 10:28 & 1998 06 02 , 10:28 & 1998 06 02 , 09:16 & Fr \\
 37   &  1998 06 24 , 10:47 & 1998 06 24 , 13:26 & 1998 06 25 , 22:33 & F+ \\
 38   &  1998 07 10 , 22:36 & 1998 07 10 , 22:36 & 1998 07 12 , 21:34 & F+ \\
 39   &  1998 08 19 , 18:40 & 1998 08 20 , 08:38 & 1998 08 21 , 20:09 & F+ \\
 40   &  1998 10 18 , 19:30 & 1998 10 19 , 04:19 & 1998 10 20 , 07:11 & F+ \\
 \hline
 
 \end{tabular}
 \end{center}
 \end{table}
 
 \begin{table}
\begin{center}
 \begin{tabular}{cccclccc}
 \hline \hline
 CME        & CME Arrival date & MC start  & MC end       & Flux rope \\
   event      & and time[UT]   &  date and   & date and     & type   \\
    number    &  (1AU)         &  time [UT]   &   time [UT]   &      \\
    \hline
41   &  1999 02 11 , 17:41 & 1999 02 11 , 17:41 & 1999 02 12 , 03:35 & Fr \\
 42   &  1999 07 02 , 00:27 & 1999 07 03 , 08:09 & 1999 07 05 , 13:13 & Fr \\
 43   &  1999 09 21 , 18:57 & 1999 09 21 , 18:57 & 1999 09 22 , 11:31 & Fr \\
 44   &  2000 02 11 , 23:34 & 2000 02 12 , 12:20 & 2000 02 13 , 00:35 & Fr \\
 45   &  2000 02 20 , 21:03 & 2000 02 21 , 14:24 & 2000 02 22 , 13:16 & Fr \\
 
 46   &  2000 03 01 , 01:58 & 2000 03 01 , 03:21 & 2000 03 02 , 03:07 & Fr \\
 47   &  2000 07 01 , 07:12 & 2000 07 01 , 07:12 & 2000 07 02 , 03:34 & Fr \\
 48   &  2000 07 11 , 22:35 & 2000 07 11 , 22:35 & 2000 07 13 , 04:33 & Fr \\
 49   &  2000 07 28 , 06:38 & 2000 07 28 , 14:24 & 2000 07 29 , 10:06 & F+ \\
 50   &  2000 09 02 , 23:16 & 2000 09 02 , 23:16 & 2000 09 03 , 22:32 & Fr \\
 51   &  2000 10 03 , 01:02 & 2000 10 03 , 09:36 & 2000 10 05 , 03:34 & F+ \\
 52   &  2000 10 12 , 22:33 & 2000 10 13 , 18:24 & 2000 10 14 , 19:12 & Fr \\
 53   &  2000 11 06 , 09:30 & 2000 11 06 , 23:05 & 2000 11 07 , 18:05 & Fr \\
 54   &  2000 11 26 , 11:43 & 2000 11 27 , 09:30 & 2000 11 28 , 09:36 & Fr \\
 55   &  2001 04 21 , 15:29 & 2001 04 22 , 00:28 & 2001 04 23 , 01:11 & Fr \\
 56   &  2001 10 21 , 16:39 & 2001 10 22 , 01:17 & 2001 10 23 , 00:47 & Fr \\
 57   &  2001 11 24 , 05:51 & 2001 11 24 , 15:47 & 2001 11 25 , 13:17 & Fr \\
 58   &  2001 12 29 , 05:16 & 2001 12 30 , 03:24 & 2001 12 30 , 19:10 & Fr \\
 59   &  2002 02 28 , 05:06 & 2002 02 28 , 19:11 & 2002 03 01 , 23:15 & Fr \\
 60   &  2002 03 18 , 13:14 & 2002 03 19 , 06:14 & 2002 03 20 , 15:36 & Fr \\
 61   &  2002 03 23 , 11:24 & 2002 03 24 , 13:11 & 2002 03 25 , 21:36 & Fr \\
 62   &  2002 04 17 , 11:01 & 2002 04 17 , 21:36 & 2002 04 19 , 08:22 & F+ \\
 63   &  2002 07 17 , 15:56 & 2002 07 18 , 13:26 & 2002 07 19 , 09:35 & Fr \\
 64   &  2002 08 18 , 18:40 & 2002 08 19 , 19:12 & 2002 08 21 , 13:25 & Fr \\
 65   &  2002 08 26 , 11:16 & 2002 08 26 , 14:23 & 2002 08 27 , 10:47 & Fr \\
 66   &  2002 09 30 , 07:54 & 2002 09 30 , 22:04 & 2002 10 01 , 20:08 & F+ \\
 67   &  2002 12 21 , 03:21 & 2002 12 21 , 10:20 & 2002 12 22 , 15:36 & Fr \\
 68   &  2003 01 26 , 21:43 & 2003 01 27 , 01:40 & 2003 01 27 , 16:04 & Fr \\
 69   &  2003 02 01 , 13:06 & 2003 02 02 , 19:11 & 2003 02 03 , 09:35 & Fr \\
 70   &  2003 03 20 , 04:30 & 2003 03 20 , 11:54 & 2003 03 20 , 22:22 & Fr \\
 71   &  2003 06 16 , 22:33 & 2003 06 16 , 17:48 & 2003 06 18 , 08:18 & Fr \\
 72   &  2003 08 04 , 20:23 & 2003 08 05 , 01:10 & 2003 08 06 , 02:23 & Fr \\
 73   &  2003 11 20 , 08:35 & 2003 11 20 , 11:31 & 2003 11 21 , 01:40 & Fr \\
 74   &  2004 04 03 , 09:55 & 2004 04 04 , 01:11 & 2004 04 05 , 19:11 & F+ \\
 75   &  2004 09 17 , 20:52 & 2004 09 18 , 12:28 & 2004 09 19 , 16:58 & Fr \\
 
 76   &  2005 05 15 , 02:10 & 2005 05 15 , 05:31 & 2005 05 16 , 22:47 & F+ \\
 77   &  2005 05 20 , 04:47 & 2005 05 20 , 09:35 & 2005 05 22 , 02:23 & F+ \\
 78   &  2005 07 17 , 14:52 & 2005 07 17 , 14:52 & 2005 07 18 , 05:59 & Fr \\
 79   &  2005 10 31 , 02:23 & 2005 10 31 , 02:23 & 2005 10 31 , 18:42 & Fr \\
 80   &  2006 02 05 , 18:14 & 2006 02 05 , 20:23 & 2006 02 06 , 11:59 & F+ \\
 81   &  2006 09 30 , 02:52 & 2006 09 30 , 08:23 & 2006 09 30 , 22:03 & F+ \\
 82   &  2006 11 18 , 07:11 & 2006 11 18 , 07:11 & 2006 11 20 , 04:47 & Fr \\
 83   &  2007 05 21 , 22:40 & 2007 05 21 , 22:45 & 2007 05 22 , 13:25 & Fr \\
 84   &  2007 06 08 , 05:45 & 2007 06 08 , 05:45 & 2007 06 09 , 05:15 & Fr \\
 85   &  2007 11 19 , 17:22 & 2007 11 20 , 00:33 & 2007 11 20 , 11:31 & Fr \\

 \hline
 
 \end{tabular}
 \end{center}
 \end{table}
 
 \begin{table}
\begin{center}
 \begin{tabular}{cccclccc}
 \hline \hline
 CME        & CME Arrival date & MC start  & MC end       & Flux rope \\
   event      & and time[UT]   &  date and   & date and     & type   \\
    number    &  (1AU)         &  time [UT]   &   time [UT]   &      \\
    \hline
 
 86   &  2008 05 23 , 01:12 & 2008 05 23 , 01:12 & 2008 05 23 , 10:46 & F+ \\
 87   &  2008 09 03 , 16:33 & 2008 09 03 , 16:33 & 2008 09 04 , 03:49 & F+ \\
 88   &  2008 09 17 , 00:43 & 2008 09 17 , 03:57 & 2008 09 18 , 08:09 & Fr \\
 89   &  2008 12 04 , 11:59 & 2008 12 04 , 16:47 & 2008 12 05 , 10:47 & Fr \\
 90   &  2008 12 17 , 03:35 & 2008 12 17 , 03:35 & 2008 12 17 , 15:35 & Fr \\
 91   &  2009 02 03 , 19:21 & 2009 02 03 , 01:12 & 2009 02 04 , 19:40 & F+ \\
 92   &  2009 03 11 , 22:04 & 2009 03 12 , 01:12 & 2009 03 13 , 01:40 & F+ \\
 93   &  2009 04 22 , 11:16 & 2009 04 22 , 14:09 & 2009 04 22 , 20:37 & Fr \\
 94   &  2009 06 03 , 13:40 & 2009 06 03 , 20:52 & 2009 06 05 , 05:31 & Fr \\
 95   &  2009 06 27 , 11:02 & 2009 06 27 , 17:59 & 2009 06 28 , 20:24 & F+ \\ 
 
 96   &  2009 07 21 , 02:53 & 2009 07 21 , 04:48 & 2009 07 22 , 03:36 & Fr \\
 97   &  2009 09 10 , 10:19 & 2009 09 10 , 10:19 & 2009 09 10 , 19:26 & Fr \\
 
 98   &  2009 09 30 , 00:44 & 2009 09 30 , 06:59 & 2009 09 30 , 19:11 & Fr \\
 99   &  2009 10 29 , 01:26 & 2009 10 29 , 01:26 & 2009 10 29 , 23:45 & F+ \\
 100   &  2009 11 14 , 10:47 & 2009 11 14 , 10:47 & 2009 11 15 , 11:45 & Fr \\
 101   &  2009 12 12 , 04:47 & 2009 12 12 , 19:26 & 2009 12 14 , 04:47 & Fr \\
 102   &  2010 01 01 , 22:04 & 2010 01 02 , 00:14 & 2010 01 03 , 09:06 & Fr \\
 103   &  2010 02 07 , 18:04 & 2010 02 07 , 19:11 & 2010 02 09 , 05:42 & Fr \\
 104*   &  2010 03 23 , 22:29 & 2010 03 23 , 22:23 & 2010 03 24 , 15:36 & Fr \\
 105*   &  2010 04 05 , 07:55 & 2010 04 05 , 11:59 & 2010 04 06 , 16:48 & Fr \\
 106*   &  2010 04 11 , 12:20 & 2010 04 11 , 21:36 & 2010 04 12 , 14:12 & Fr \\
 107   &  2010 05 28 , 01:55 & 2010 05 29 , 19:12 & 2010 05 29 , 17:58 & Fr \\
 108*   &  2010 06 21 , 03:35 & 2010 06 21 , 06:28 & 2010 06 22 , 12:43 & Fr \\
 109*   &  2010 09 15 , 02:24 & 2010 09 15 , 02:24 & 2010 09 16 , 11:58 & Fr \\
 110*   &  2010 10 31 , 02:09 & 2010 10 30 , 05:16 & 2010 11 01 , 20:38 & Fr \\
 111   &  2010 12 19 , 00:35 & 2010 12 19 , 22:33 & 2010 12 20 , 22:14 & F+ \\
 112   &  2011 01 24 , 06:43 & 2011 01 24 , 10:33 & 2011 01 25 , 22:04 & F+ \\
 113*   &  2011 03 29 , 15:12 & 2011 03 29 , 23:59 & 2011 04 01 , 14:52 & Fr \\
 114   &  2011 05 28 , 00:14 & 2011 05 28 , 05:31 & 2011 05 28 , 22:47 & F+ \\
 115   &  2011 06 04 , 20:06 & 2011 06 05 , 01:12 & 2011 06 05 , 18:13 & Fr \\
 116   &  2011 07 03 , 19:12 & 2011 07 03 , 19:12 & 2011 07 04 , 19:12 & Fr \\
 117*   &  2011 09 17 , 02:57 & 2011 09 17 , 15:35 & 2011 09 18 , 21:07 & Fr \\
 118   &  2012 02 14 , 07:11 & 2012 02 14 , 20:52 & 2012 02 16 , 04:47 & Fr \\
 119   &  2012 04 05 , 14:23 & 2012 04 05 , 19:41 & 2012 04 06 , 21:36 & Fr \\
 120   &  2012 05 03 , 00:59 & 2012 05 04 , 03:36 & 2012 05 05 , 11:22 & Fr \\
 121   &  2012 05 16 , 12:28 & 2012 05 16 , 16:04 & 2012 05 18 , 02:11 & Fr \\
 122   &  2012 06 11 , 02:52 & 2012 06 11 , 11:31 & 2012 06 12 , 05:16 & Fr \\
 123*   &  2012 06 16 , 09:03 & 2012 06 16 , 22:01 & 2012 06 17 , 11:23 & F+ \\
 124*   &  2012 07 14 , 17:39 & 2012 07 15 , 06:14 & 2012 07 17 , 03:22 & Fr \\
 125   &  2012 08 12 , 12:37 & 2012 08 12 , 19:12 & 2012 08 13 , 05:01 & Fr \\
 126   &  2012 08 18 , 03:25 & 2012 08 18 , 19:12 & 2012 08 19 , 08:22 & Fr \\
 127*   &  2012 10 08 , 04:12 & 2012 10 08 , 15:50 & 2012 10 09 , 17:17 & Fr \\
 128   &  2012 10 12 , 08:09 & 2012 10 12 , 18:09 & 2012 10 13 , 09:14 & Fr \\
 129*   &  2012 10 31 , 14:28 & 2012 10 31 , 23:35 & 2012 11 02 , 05:21 & F+ \\
 130*   &  2013 03 17 , 05:21 & 2013 03 17 , 14:09 & 2013 03 19 , 16:04 & Fr \\

 \hline
 
 \end{tabular}
 \end{center}
 \end{table}
 
  \begin{table}
\begin{center}
 \begin{tabular}{cccclccc}
 \hline\hline
 CME        & CME Arrival date & MC start  & MC end       & Flux rope \\
   event      & and time[UT]   &  date and   & date and     & type   \\
    number    &  (1AU)         &  time [UT]   &   time [UT]   &      \\
    \hline
 131*   &  2013 04 13 , 22:13 & 2013 04 14 , 17:02 & 2013 04 17 , 05:30 & F+ \\
 132   &  2013 04 30 , 08:52 & 2013 04 30 , 12:00 & 2013 05 01 , 07:12 & Fr \\
 133   &  2013 05 14 , 02:23 & 2013 05 14 , 06:00 & 2013 05 15 , 06:28 & Fr \\
 134   &  2013 06 06 , 02:09 & 2013 06 06 , 14:23 & 2013 06 08 , 00:00 & F+ \\
 135   &  2013 06 27 , 13:51 & 2013 06 28 , 02:23 & 2013 06 29 , 11:59 & Fr \\
 136   &  2013 09 01 , 06:14 & 2013 09 01 , 13:55 & 2013 09 02 , 01:56 & Fr \\
 137   &  2013 10 30 , 18:14 & 2013 10 30 , 18:14 & 2013 10 31 , 05:30 & Fr \\
 138   &  2013 11 08 , 21:07 & 2013 11 08 , 23:59 & 2013 11 09 , 06:14 & Fr \\
 139   &  2013 11 23 , 00:14 & 2013 11 23 , 04:47 & 2013 11 23 , 15:35 & Fr \\
 140   &  2013 12 14 , 16:47 & 2013 12 15 , 16:47 & 2013 12 16 , 05:30 & Fr \\
 141   &  2013 12 24 , 20:36 & 2013 12 25 , 04:47 & 2013 12 25 , 17:59 & F+ \\
 142   &  2014 04 05 , 09:58 & 2014 04 05 , 22:18 & 2014 04 07 , 14:24 & Fr \\
 143   &  2014 04 11 , 06:57 & 2014 04 11 , 06:57 & 2014 04 12 , 20:52 & F+ \\
 144   &  2014 04 20, 10:20 & 2014 04 21 , 07:41 & 2014 04 22 , 06:12 & Fr \\
 145   &  2014 04 29 , 19:11 & 2014 04 29 , 19:11 & 2014 04 30 , 16:33 & Fr \\
 146   &  2014 06 29 , 04:47 & 2014 06 29 , 20:53 & 2014 06 30 , 11:15 & Fr \\
 147   &  2014 08 19 , 05:49 & 2014 08 19 , 17:59 & 2014 08 21 , 19:09 & F+ \\
 148   &  2014 08 26 , 02:40 & 2014 08 27 , 03:07 & 2014 08 27 , 21:49 & Fr \\
 149   &  2015 01 07 , 05:38 & 2015 01 07 , 06:28 & 2015 01 07 , 21:07 & F+ \\
 150   &  2015 09 07 , 13:05 & 2015 09 07 , 23:31 & 2015 09 09 , 14:52 & F+ \\
 151   &  2015 10 06 , 21:35 & 2015 10 06 , 21:35 & 2015 10 07 , 10:03 & Fr \\
 152   &  2015 12 19 , 15:35 & 2015 12 20 , 13:40 & 2015 12 21 , 23:02 & Fr \\
 \hline
 
  \end{tabular}
  \end{center}
\end{table}   

\begin{landscape}
\begin{table}
\captionof{table}{Column 1 gives the event number corresponding to the events listed in Table~\ref{S - Table A}. We note that some events in Table~\ref{S - Table A} do not appear here. Column 2 gives the time duration of the sheath in days. We note that the sheath duration can be zero for some events. Column 3 gives the time duration of the MC in days. Coulmns 4 to 12 list the quantities defined in Eq~\ref{eqP} for each event.}\label{Table_comp}
    \centering
    \begin{tabular}{llllllllllll}
    \hline
        Event & Sh dur & MC dur & $P_{\rm B sh}$ & $P_{\rm V sh}$ & $P_{\rm d  sh}$ & $P_{\rm B MC}$ & $P_{\rm V MC}$ & $P_{\rm d MC}$ & $P_{\rm B bg}$ & $P_{\rm V bg}$ & $P_{\rm d bg}$ \\ \hline
        1 & 0.449 & 0.655 & 1.64E-02 & 6.62E+00 & 4.95E-01 & 2.05E-02 & 1.17E+00 & 7.90E-02 & 2.40E-03 & 6.60E+00 & 4.24E-03 \\ \hline
        2 & 0.251 & 1.03 & 2.52E-02 & 3.80E+00 & 7.79E-03 & 7.04E-03 & 4.52E+00 & 3.14E-03 & 3.14E-03 & 1.43E+01 & 3.18E-01 \\ \hline
        3 & 0.21 & 2.1 & 9.47E-02 & 7.32E-01 & 1.55E+00 & 4.30E-03 & 5.59E+00 & 4.37E-02 & 9.42E-03 & 1.14E+00 & 2.63E-01 \\ \hline
        4 & 0.39 & 0.85 & 3.85E-02 & 1.72E+00 & 1.39E-01 & 9.14E-03 & 1.01E+00 & 1.30E-01 & 2.76E-03 & 2.19E+00 & 3.16E-01 \\ \hline
        5 & 0.485 & 0.74 & 1.84E-02 & 1.44E+00 & 4.19E-02 & 9.52E-02 & 6.43E+00 & 1.41E-01 & 1.72E-03 & 3.31E+00 & 3.70E-01 \\ \hline
        6 & 0.354 & 1.3 & 1.78E-01 & 1.52E+00 & 2.54E+00 & 1.96E-02 & 7.21E-01 & 8.49E-01 & 2.74E-03 & 9.51E-01 & 4.82E-02 \\ \hline
        7 & 0 & 0.74 & - & - & - & 2.19E-02 & 1.76E+00 & 4.15E-01 & 1.40E-03 & 5.80E+00 & 2.96E-02 \\ \hline
        8 & 0 & 0.4 & - & - & - & 4.13E-02 & 1.81E+00 & 1.71E-01 & 4.82E-03 & 3.43E-01 & 1.25E-01 \\ \hline
        9 & 0.117 & 0.43 & 1.05E-01 & 1.23E+02 & 1.17E-01 & 7.94E-03 & 9.64E-01 & 9.59E-03 & 6.20E-03 & 4.45E+00 & 8.23E-03 \\ \hline
        10 & 0 & 1.48 & - & - & - & 8.80E-03 & 1.58E+00 & 1.89E-01 & 4.60E-03 & 4.41E+00 & 3.34E-01 \\ \hline
        11 & 0.174 & 0.7 & 4.57E-02 & 1.72E+00 & 3.82E-01 & 3.15E-02 & 6.27E-01 & 9.27E-02 & 6.33E-03 & 1.20E+00 & 7.91E-02 \\ \hline
        12 & 0.15 & 1.05 & 1.82E-02 & 2.98E-01 & 4.77E-02 & 6.00E-03 & 3.04E-01 & 9.40E-03 & 5.95E-04 & 1.20E+00 & 7.82E-02 \\ \hline
        13 & 0.07 & 1.36 & 1.14E-01 & 4.71E+00 & 1.66E+00 & 1.11E-02 & 2.15E+00 & 7.66E-02 & 2.66E-03 & 8.48E-01 & 2.22E-01 \\ \hline
        14 & 0.163 & 0.93 & 2.76E-02 & 9.31E+00 & 5.91E-02 & 1.24E-02 & 2.54E+00 & 6.00E+00 & 5.12E-03 & 6.93E-01 & 9.28E-02 \\ \hline
        15 & 0.53 & 0.56 & 9.51E-02 & 5.51E+00 & 5.04E-01 & 1.49E-01 & 1.18E+00 & 2.53E-01 & 4.84E-03 & 4.74E-01 & 1.81E-02 \\ \hline
        16 & 0.075 & 1.8 & 1.16E-01 & 2.48E+00 & 4.21E-01 & 6.33E-03 & 2.31E+00 & 4.99E-02 & 4.38E-04 & 1.87E+00 & 7.71E-04 \\ \hline
        17 & 0.364 & 0.69 & 7.66E-02 & 2.33E+00 & 1.05E+00 & 1.87E-01 & 6.05E+00 & 6.11E-02 & 1.58E-03 & 5.32E-01 & 1.20E-02 \\ \hline
        18 & 0.268 & 1.35 & 1.37E-01 & 1.07E+01 & 8.06E-01 & 8.15E-03 & 9.75E-01 & 5.36E-02 & 4.26E-03 & 1.02E-01 & 3.11E-02 \\ \hline
        19 & 0.6 & 0.696 & 1.62E-02 & 7.03E-01 & 1.13E-01 & 1.44E-02 & 1.71E+00 & 1.74E-01 & 6.03E-04 & 1.46E+00 & 2.51E-01 \\ \hline
        20 & 0.23 & 1.7 & 2.33E-02 & 2.20E+00 & 2.11E-01 & 2.40E-03 & 1.99E+00 & 9.56E-03 & 2.20E-03 & 1.25E+00 & 2.18E-01 \\ \hline
        21 & 0.151 & 1.18 & 1.45E-01 & 8.81E+00 & 3.42E+00 & 2.15E-02 & 5.68E-01 & 5.59E-02 & 8.43E-04 & 3.77E-01 & 1.17E-01 \\ \hline
        22 & 0.156 & 0.52 & 3.97E-01 & 5.37E+00 & 4.03E+00 & 1.74E-02 & 1.10E+01 & 3.45E-01 & 6.59E-03 & 1.12E+00 & 3.87E-02 \\ \hline
        23 & 0.192 & 0.57 & 2.48E-02 & 1.54E+01 & 2.89E-02 & 1.32E-02 & 1.06E+00 & 2.72E-02 & 3.15E-03 & 3.37E+00 & 1.45E-01 \\ \hline
        24 & 0.41 & 0.514 & 1.17E-02 & 3.35E+00 & 2.69E-01 & 3.44E-02 & 2.81E+00 & 2.51E-01 & 2.62E-03 & 2.43E+00 & 1.31E-02 \\ \hline
        25 & 0.15 & 1.83 & 1.81E-01 & 1.12E+00 & 3.60E+00 & 2.30E-03 & 1.58E+00 & 1.75E-02 & 4.20E-03 & 6.50E-01 & 6.63E-03 \\ \hline
        26 & 0.224 & 1.25 & 7.43E-02 & 9.18E-01 & 1.16E-01 & 1.32E-03 & 1.61E+00 & 2.51E-02 & 7.06E-03 & 2.06E+00 & 4.40E-01 \\ \hline
        27 & 0.517 & 1.27 & 6.96E-03 & 6.88E+00 & 4.56E-02 & 1.05E-02 & 2.59E+00 & 2.92E-01 & 3.77E-03 & 6.63E-01 & 2.40E-01 \\ \hline
        28 & 0.316 & 1.7 & 4.43E-02 & 5.35E+00 & 5.80E-01 & 2.86E-02 & 2.60E+00 & 1.88E-01 & 1.28E-03 & 1.22E-01 & 3.00E-02 \\ \hline
        29 & 0.346 & 1.05 & 3.00E-01 & 9.94E+00 & 3.72E+00 & 1.18E-01 & 3.67E+00 & 1.86E-01 & 3.07E-03 & 7.46E-01 & 1.33E-01 \\ \hline
        30 & 0.35 & 0.97 & 7.36E-02 & 4.12E-01 & 1.42E+00 & 1.65E-02 & 3.42E+00 & 1.86E-01 & 1.47E-03 & 2.73E-01 & 5.78E-03 \\ \hline
        31 & 0.537 & 1.23 & 6.42E-02 & 2.10E+00 & 2.68E-01 & 1.56E-02 & 2.31E+00 & 4.71E-02 & 2.63E-04 & 1.46E+00 & 1.74E-01 \\ \hline
        32 & 0.88 & 1.45 & 5.14E-03 & 4.24E+00 & 2.31E-01 & 1.69E-03 & 8.29E-01 & 1.69E-02 & 6.91E-03 & 3.76E+00 & 3.21E-01 \\ \hline
        33 & 0.15 & 0.77 & 6.37E-02 & 7.28E-01 & 2.30E-03 & 6.92E-03 & 6.41E-01 & 1.30E-01 & 2.23E-03 & 1.01E+01 & 3.05E-01 \\ \hline
        34 & 0.2 & 1.18 & 8.59E-02 & 2.32E+00 & 1.86E+00 & 1.70E-03 & 1.82E+00 & 7.64E-02 & 2.38E-03 & 3.85E+00 & 2.10E-01 \\ \hline
        35 & 0.59 & 1.22 & 5.66E-02 & 1.81E+01 & 2.73E-01 & 5.07E-02 & 1.96E+01 & 8.30E-01 & 2.76E-03 & 4.34E+00 & 3.59E-01 \\ \hline
        37 & 0.11 & 1.38 & 1.03E-01 & 7.30E+00 & 2.21E-02 & 2.08E-02 & 4.14E+00 & 1.03E-01 & 1.42E-03 & 8.67E-01 & 2.02E-03 \\ \hline
        38 & 0 & 1.96 & - & - & - & 1.41E-02 & 1.72E+00 & 2.46E-02 & 2.35E-03 & 8.91E-01 & 2.53E-02 \\ \hline
        39 & 0.582 & 1.48 & 8.45E-02 & 2.25E+00 & 5.74E-01 & 2.60E-02 & 1.16E+00 & 2.69E-02 & 7.70E-04 & 1.83E+00 & 1.88E-01 \\ \hline
        40 & 0.367 & 1.12 & 3.61E-01 & 9.25E+00 & 3.01E+00 & 8.71E-02 & 1.75E+00 & 1.52E-01 & 2.51E-03 & 1.05E+00 & 3.54E-02 \\ \hline
\hline
\end{tabular}
\end{table}

\begin{table}
    \begin{tabular}{llllllllllll}
    \hline
        Event & Sh dur & MC dur & $P_{\rm B sh}$ & $P_{\rm V sh}$ & $P_{\rm d  sh}$ & $P_{\rm B MC}$ & $P_{\rm V MC}$ & $P_{\rm d MC}$ & $P_{\rm B bg}$ & $P_{\rm V bg}$ & $P_{\rm d bg}$ \\ \hline        
        41 & 0 & 0.41 & - & - & - & 1.54E-01 & 9.15E+00 & 8.85E+00 & 1.66E-03 & 5.39E-01 & 1.11E-01 \\ \hline
        42 & 1.32 & 2.21 & 1.57E-02 & 5.01E+00 & 1.81E-02 & 1.94E-04 & 7.26E+00 & 2.38E-02 & 1.88E-03 & 1.36E+01 & 2.01E-01 \\ \hline
        43 & 0 & 0.69 & - & - & - & 1.18E-02 & 1.80E-01 & 5.79E-02 & 5.31E-03 & 4.91E-01 & 8.08E-02 \\ \hline
        44 & 0.532 & 0.51 & 1.62E-01 & 8.14E+00 & 4.80E-01 & 6.12E-02 & 1.56E+00 & 1.75E-01 & 1.22E-03 & 8.64E+00 & 7.12E-04 \\ \hline
        45 & 0.723 & 0.953 & 7.21E-02 & 3.14E+00 & 6.75E-01 & 1.56E-02 & 3.22E+00 & 1.99E-01 & 3.86E-04 & 1.10E+00 & 1.11E-01 \\ \hline
        46 & 0.06 & 0.99 & 6.19E-02 & 9.35E-01 & 7.93E-02 & 8.21E-03 & 6.22E+00 & 1.02E-02 & 1.25E-03 & 2.98E+00 & 2.23E-01 \\ \hline
        47 & 0 & 0.85 & - & - & - & 2.97E-03 & 1.32E+00 & 6.65E-02 & 1.04E-03 & 1.96E+00 & 2.09E-03 \\ \hline
        48 & 0 & 1.25 & - & - & - & 4.41E-03 & 3.36E+00 & 1.02E-01 & 3.71E-03 & 1.50E+00 & 3.40E-03 \\ \hline
        49 & 0.323 & 0.82 & 8.18E-02 & 1.42E+01 & 1.19E+00 & 2.61E-02 & 8.12E-01 & 3.00E-01 & 1.00E-02 & 4.66E+00 & 2.55E-01 \\ \hline
        50 & 0 & 0.97 & - & - & - & 9.19E-03 & 2.74E+00 & 1.16E-02 & 4.90E-04 & 2.95E+00 & 1.66E-01 \\ \hline
        51 & 0.357 & 1.75 & 3.83E-02 & 2.02E+00 & 2.50E-01 & 2.50E-02 & 1.27E+00 & 8.49E-02 & 2.70E-03 & 8.07E+00 & 1.45E-01 \\ \hline
        52 & 0.827 & 1.03 & 4.13E-02 & 3.07E+00 & 1.14E+00 & 3.07E-03 & 6.09E-01 & 4.54E-02 & 3.98E-03 & 1.96E-01 & 4.18E-02 \\ \hline
        53 & 0.566 & 0.792 & 4.86E-02 & 5.50E+00 & 9.85E-01 & 6.00E-02 & 9.84E+00 & 4.68E-02 & 1.12E-03 & 7.94E-01 & 1.02E+00 \\ \hline
        54 & 0.9 & 1.0 & 1.25E-01 & 4.09E+00 & 4.54E-01 & 4.14E-02 & 7.29E+00 & 1.75E-01 & 6.13E-03 & 1.95E+00 & 1.54E-01 \\ \hline
        55 & 0.374 & 1.03 & 2.16E-02 & 8.09E-01 & 5.05E-01 & 2.77E-02 & 1.91E+00 & 7.16E-02 & 2.03E-03 & 2.36E+00 & 1.66E-01 \\ \hline
        56 & 0.36 & 0.98 & 2.01E-01 & 9.54E+00 & 1.87E+01 & 3.10E-02 & 1.91E+01 & 2.79E-01 & 2.14E-03 & 2.12E+00 & 8.51E-01 \\ \hline
        57 & 0.414 & 0.89 & 4.37E+00 & 3.59E+03 & 3.20E+00 & 5.15E-02 & 5.35E+01 & 1.64E-02 & 3.01E-03 & 7.12E-01 & 1.03E-01 \\ \hline
        58 & 0.92 & 0.657 & 1.02E-01 & 6.74E+00 & 7.76E-01 & 4.36E-02 & 4.27E+00 & 8.55E-01 & 1.49E-03 & 7.97E-01 & 7.31E-02 \\ \hline
        59 & 0.587 & 1.17 & 9.42E-02 & 2.44E+00 & 2.31E-01 & 2.04E-02 & 6.27E-01 & 1.23E-01 & 4.52E-04 & 7.30E-01 & 6.91E-02 \\ \hline
        60 & 0.7 & 1.39 & 8.38E-02 & 4.64E+00 & 1.06E+00 & 2.68E-02 & 1.32E+01 & 4.47E-02 & 1.68E-03 & 1.62E+00 & 1.96E-01 \\ \hline
        61 & 1.07 & 1.35 & 4.51E-03 & 2.89E+00 & 9.95E-02 & 1.73E-02 & 1.19E+00 & 9.60E-02 & 5.50E-03 & 1.65E+00 & 8.33E-01 \\ \hline
        62 & 0.44 & 1.45 & 9.52E-01 & 2.65E+01 & 1.58E+00 & 8.16E-03 & 8.34E+00 & 5.26E-02 & 6.58E-03 & 3.69E-01 & 2.12E-01 \\ \hline
        63 & 0.89 & 0.84 & 6.29E-02 & 5.40E+00 & 9.11E-02 & 1.83E-03 & 4.10E+01 & 4.43E-03 & 7.99E-04 & 2.72E+00 & 9.66E-02 \\ \hline
        64 & 1.02 & 1.76 & 2.36E-02 & 5.62E+00 & 6.84E-02 & 4.49E-03 & 2.57E+00 & 5.13E-03 & 7.41E-04 & 6.73E+00 & 5.14E-01 \\ \hline
        66 & 0.59 & 0.92 & 7.49E-02 & 8.94E-01 & 9.36E-01 & 1.05E-01 & 2.11E+00 & 1.96E-01 & 3.16E-03 & 5.06E+00 & 7.03E-01 \\ \hline
        67 & 0.291 & 1.22 & 4.07E-02 & 1.20E+00 & 7.39E-01 & 5.98E-02 & 3.59E+00 & 8.24E-02 & 6.31E-04 & 3.73E+00 & 4.10E-01 \\ \hline
        68 & 0.16 & 0.6 & 3.09E-02 & 1.43E+00 & 1.36E-01 & 5.40E-03 & 8.62E+00 & 5.11E-02 & 1.68E-03 & 5.94E+00 & 4.06E-01 \\ \hline
        69 & 1.25 & 0.6 & 5.22E-03 & 5.39E+01 & 2.47E-01 & 2.97E-02 & 1.27E+00 & 2.04E-02 & 6.58E-03 & 1.45E+01 & 3.42E+00 \\ \hline
        70 & 0.31 & 0.43 & 6.01E-02 & 2.13E+01 & 4.42E-03 & 9.29E-03 & 2.35E+01 & 5.49E-02 & 1.92E-03 & 4.70E+00 & 5.74E-01 \\ \hline
        71 & 0.8 & 0.6 & 5.28E-03 & 2.49E+00 & 6.53E-02 & 1.32E-01 & 3.50E+00 & 1.20E-01 & 1.81E-03 & 3.66E+00 & 2.77E-01 \\ \hline
        72 & 0.19 & 1.05 & 1.55E-02 & 6.64E+00 & 1.13E-01 & 1.69E-02 & 1.65E+00 & 1.80E-01 & 9.74E-04 & 7.56E+00 & 4.25E-01 \\ \hline
        73 & 0.12 & 0.59 & 4.20E-01 & 8.03E+01 & 2.29E+00 & 1.69E+00 & 4.05E+01 & 2.33E+00 & 2.65E-03 & 2.84E+00 & 6.40E-01 \\ \hline
        76 & 0.139 & 1.72 & 2.70E+00 & 1.28E+01 & 2.12E-01 & 5.61E-01 & 5.31E+01 & 3.62E-02 & 1.66E-03 & 2.04E+00 & 3.79E-01 \\ \hline
        77 & 0.2 & 1.7 & 8.11E-03 & 3.76E-01 & 4.37E+00 & 3.41E-02 & 3.06E+00 & 2.44E-02 & 7.77E-03 & 6.41E+00 & 1.12E+00 \\ \hline
        78 & 0 & 0.63 & - & - & - & 5.82E-02 & 3.19E+00 & 3.71E-01 & 1.88E-03 & 4.24E+00 & 5.17E-01 \\ \hline
        79 & 0 & 0.68 & - & - & - & 1.39E-02 & 1.38E+00 & 1.25E-01 & 4.86E-04 & 5.20E-01 & 8.07E-02 \\ \hline
        80 & 0.09 & 0.65 & 1.03E-01 & 2.34E+00 & 2.04E+00 & 2.10E-02 & 1.08E+00 & 8.60E-02 & 1.63E-03 & 8.75E-01 & 2.98E-01 \\ \hline
        81 & 0.23 & 0.569 & 4.21E-02 & 1.66E+01 & 4.48E+00 & 3.86E-02 & 4.31E+00 & 4.04E-01 & 2.51E-04 & 8.45E+00 & 1.30E-01 \\ \hline
        83 & 0.003 & 0.61 & 4.30E+00 & 8.63E+01 & 2.62E+01 & 4.11E-02 & 5.99E-01 & 5.94E-02 & 2.75E-03 & 6.54E+00 & 2.32E-01 \\ \hline
        84 & 0 & 0.98 & - & - & - & 6.93E-03 & 3.78E-01 & 1.59E-02 & 3.57E-04 & 2.49E-01 & 3.27E-02 \\ \hline
\hline
 \end{tabular}
 \end{table}

\begin{table}[!ht]
    \centering
    \begin{tabular}{llllllllllll}
    \hline
        Event & Sh dur & MC dur & $P_{\rm B sh}$ & $P_{\rm V  sh}$ & $P_{\rm d  sh}$ & $P_{\rm B  MC}$ & $P_{\rm V  MC}$ & $P_{\rm d  MC}$ & $P_{\rm B  bg}$ & $P_{\rm V  bg}$ & $P_{\rm d  bg}$ \\ \hline   
        85 & 0.3 & 0.457 & 1.08E-01 & 1.01E+01 & 2.67E+00 & 2.63E-02 & 2.01E+00 & 1.23E+00 & 6.11E-04 & 3.36E+00 & 8.16E-02 \\ \hline
        86 & 0 & 0.4 & - & - & - & 8.81E-04 & 2.89E+00 & 2.83E-01 & 7.55E-04 & 1.39E+00 & 1.89E-03 \\ \hline
        87 & 0 & 0.47 & - & - & - & 1.42E-02 & 5.05E+00 & 2.15E-01 & 2.12E-03 & 1.58E-01 & 4.00E-02 \\ \hline
        88 & 0.135 & 1.17 & 8.35E-03 & 7.61E+00 & 1.25E-02 & 2.95E-03 & 2.82E+00 & 2.61E-02 & 6.98E-04 & 2.65E+00 & 2.22E-01 \\ \hline
        89 & 0.2 & 0.75 & 2.50E-02 & 2.11E+00 & 2.67E-02 & 1.23E-03 & 2.50E+00 & 7.29E-02 & 5.75E-04 & 3.51E+00 & 3.65E-02 \\ \hline
        90 & 0 & 0.5 & - & - & - & 1.94E-02 & 1.56E+00 & 5.44E-01 & 1.79E-03 & 3.72E-01 & 3.62E-02 \\ \hline
        91 & 0.243 & 0.77 & 5.24E-02 & 4.83E+00 & 1.27E+00 & 1.56E-02 & 1.28E+00 & 3.80E-01 & 2.64E-04 & 1.82E+00 & 2.16E-01 \\ \hline
        92 & 0.13 & 1.02 & 8.02E-02 & 8.00E+00 & 1.98E+00 & 5.51E-02 & 1.48E+00 & 2.91E-01 & 1.67E-03 & 3.84E+00 & 6.61E-02 \\ \hline
        93 & 0.12 & 0.27 & 3.87E-03 & 5.84E-01 & 5.22E-04 & 1.54E-03 & 2.31E+00 & 1.74E-03 & 1.55E-03 & 8.46E-01 & 1.25E-03 \\ \hline
        94 & 0.3 & 1.36 & 1.31E-02 & 3.08E-01 & 4.30E-01 & 3.05E-03 & 1.80E+00 & 3.68E-02 & 8.79E-04 & 3.09E-01 & 6.24E-02 \\ \hline
        95 & 0.29 & 1.1 & 6.06E-03 & 6.61E+00 & 5.78E-02 & 6.83E-03 & 1.89E+00 & 1.22E-01 & 1.13E-03 & 2.90E+00 & 2.53E-01 \\ \hline
        96 & 0.08 & 0.95 & 8.41E-02 & 2.07E+00 & 1.34E-01 & 2.47E-02 & 1.51E+00 & 3.18E-01 & 4.23E-04 & 8.17E-01 & 8.89E-03 \\ \hline
        97 & 0 & 0.38 & - & - & - & 1.36E-02 & 1.84E-01 & 7.13E-02 & 8.91E-04 & 4.38E-01 & 1.07E-01 \\ \hline
        98 & 0.26 & 0.51 & 5.73E-02 & 1.15E+00 & 3.50E-01 & 1.41E-02 & 7.51E-01 & 4.80E-02 & 1.20E-03 & 6.50E-01 & 1.61E-01 \\ \hline
        100 & 0 & 1.04 & - & - & - & 2.29E-03 & 1.58E+00 & 4.50E-02 & 9.91E-04 & 1.06E-01 & 6.61E-02 \\ \hline
        101 & 0.61 & 1.39 & 6.00E-03 & 3.38E-01 & 5.20E-02 & 3.38E-03 & 9.43E-01 & 8.39E-02 & 6.84E-04 & 1.21E-01 & 3.48E-02 \\ \hline
        102 & 0.09 & 1.37 & 1.83E-03 & 9.08E-03 & 7.65E-02 & 1.61E-03 & 4.24E-01 & 1.01E-01 & 1.53E-03 & 3.27E+00 & 1.09E-01 \\ \hline
        103 & 0.046 & 1.44 & 2.06E-02 & 0.00E+00 & 4.88E-02 & 1.07E-02 & 1.93E+00 & 5.80E-02 & 1.46E-04 & 8.83E-01 & 8.51E-03 \\ \hline
        104 & 0.003 & 0.71 & 1.12E+00 & 3.02E+02 & 5.07E+00 & 3.99E-03 & 1.36E+00 & 7.35E-02 & 6.31E-04 & 1.22E+00 & 3.63E-03 \\ \hline
        105 & 0.17 & 1.2 & 1.79E-01 & 1.82E+00 & 1.41E-01 & 2.45E-02 & 2.58E+01 & 3.17E-02 & 1.55E-03 & 7.33E+00 & 2.57E-02 \\ \hline
        106 & 0.386 & 0.69 & 1.04E-02 & 3.42E+00 & 9.87E-03 & 1.53E-02 & 1.55E+00 & 1.29E-01 & 2.53E-03 & 5.83E-01 & 5.22E-02 \\ \hline
        107 & 0.72 & 0.95 & 1.48E-02 & 7.81E-01 & 1.35E-01 & 4.82E-03 & 9.57E-01 & 4.97E-02 & 5.79E-03 & 1.68E+00 & 6.39E-01 \\ \hline
        108 & 0.12 & 1.26 & 1.19E-02 & 3.23E-01 & 3.14E-01 & 2.03E-03 & 9.94E-01 & 3.88E-02 & 6.14E-04 & 1.70E+00 & 1.78E-03 \\ \hline
        109 & 0 & 1.4 & - & - & - & 9.86E-03 & 1.25E+00 & 2.41E-02 & 6.57E-05 & 2.70E+00 & 6.60E-04 \\ \hline
        110 & 0.13 & 1.64 & 3.06E-02 & 5.88E+00 & 1.86E-01 & 4.21E-03 & 1.04E+00 & 1.02E-01 & 1.54E-03 & 8.28E+00 & 3.11E-01 \\ \hline
        111 & 0.08 & 0.9 & 8.73E-03 & 2.78E+00 & 6.63E-01 & 3.62E-03 & 1.34E+00 & 1.65E-01 & 3.29E-04 & 1.22E+00 & 6.13E-02 \\ \hline
        112 & 0.16 & 1.48 & 1.93E-02 & 7.65E+00 & 1.94E-01 & 3.49E-03 & 1.82E+00 & 9.77E-02 & 3.75E-04 & 2.85E+00 & 5.98E-04 \\ \hline
        113 & 0.36 & 2.6 & 3.54E-02 & 1.34E+00 & 1.87E+00 & 3.62E-03 & 1.21E+00 & 2.39E-02 & 5.21E-04 & 1.06E+00 & 4.99E-02 \\ \hline
        114 & 0.22 & 0.72 & 1.23E-02 & 6.05E+00 & 8.92E-03 & 6.10E-03 & 5.33E+00 & 7.09E-02 & 1.69E-03 & 6.09E-01 & 1.20E-01 \\ \hline
        115 & 0.21 & 0.71 & 3.25E-01 & 2.01E+01 & 3.13E+00 & 2.24E-01 & 4.65E+00 & 2.13E+00 & 7.78E-03 & 2.00E+00 & 1.21E-01 \\ \hline
        116 & 0 & 1 & - & - & - & 2.36E-03 & 2.05E+00 & 1.78E-03 & 9.42E-04 & 3.03E+00 & 1.09E-01 \\ \hline
        117 & 0.526 & 1.23 & 2.69E-02 & 1.77E+01 & 5.43E-01 & 4.28E-02 & 1.46E+00 & 1.11E-02 & 2.84E-03 & 3.31E+00 & 1.46E-01 \\ \hline
        118 & 0.57 & 1.33 & 4.22E-03 & 1.08E+00 & 6.74E-02 & 1.99E-03 & 7.42E-01 & 1.08E-02 & 3.31E-03 & 9.76E-01 & 6.52E-02 \\ \hline
        119 & 0.22 & 1.08 & 2.59E-02 & 0.00E+00 & 8.63E-01 & 1.14E-02 & 1.50E+00 & 6.66E-01 & 4.20E-04 & 1.01E+00 & 6.47E-02 \\ \hline
        120 & 1.11 & 1.3 & 1.57E-02 & 2.90E-01 & 7.55E-02 & 5.95E-03 & 5.06E-01 & 9.14E-03 & 9.86E-04 & 1.45E+00 & 1.48E-03 \\ \hline
        121 & 0.15 & 1.42 & 3.70E-03 & 1.99E+00 & 2.94E-01 & 9.80E-03 & 1.39E+00 & 1.96E-02 & 3.31E-04 & 1.46E+00 & 1.11E-01 \\ \hline
        122 & 0.36 & 0.74 & 8.65E-03 & 2.32E+00 & 3.91E-02 & 1.35E-02 & 1.22E+00 & 1.56E-02 & 7.21E-04 & 1.77E+00 & 5.95E-04 \\ \hline
        124 & 0.524 & 1.89 & 1.86E-01 & 2.36E+01 & 2.33E-01 & 1.06E-01 & 1.38E+01 & 1.50E-02 & 5.85E-03 & 4.74E-01 & 6.09E-02 \\ \hline
        125 & 0.274 & 0.41 & 5.66E-02 & 8.34E-01 & 1.16E-01 & 2.04E-02 & 5.23E-01 & 3.97E-02 & 1.40E-03 & 1.17E+00 & 1.27E-03 \\ \hline
\hline
 \end{tabular}
 \end{table}

\begin{table}[!ht]
    \centering
    \begin{tabular}{llllllllllll}
    \hline
        Ev & Sh dur & MC dur & $P_{\rm B sh}$ & $P_{\rm V sh}$ & $P_{\rm d  sh}$ & $P_{\rm B MC}$ & $P_{\rm V  MC}$ & $P_{\rm d  MC}$ & $P_{\rm B  bg}$ & $P_{\rm V  bg}$ & $P_{\rm d  bg}$ \\ \hline

        126 & 0.66 & 0.55 & 7.30E-03 & 5.05E-01 & 1.62E-02 & 8.21E-03 & 2.69E+00 & 1.00E-01 & 4.08E-03 & 4.99E-01 & 4.99E-02 \\ \hline
        127 & 0.485 & 1.06 & 1.10E-01 & 5.72E+00 & 9.06E-02 & 5.76E-03 & 4.23E+00 & 2.97E-02 & 2.39E-03 & 2.81E+00 & 1.35E-01 \\ \hline
        128 & 0.43 & 0.614 & 2.74E-03 & 6.34E+00 & 2.85E-02 & 2.41E-02 & 2.65E+00 & 5.06E-02 & 3.02E-04 & 1.04E+00 & 6.77E-02 \\ \hline
        129 & 0.38 & 1.24 & 2.98E-02 & 1.07E+00 & 3.41E-01 & 2.71E-02 & 1.78E+00 & 1.89E-01 & 4.60E-03 & 2.92E+00 & 2.13E-01 \\ \hline
        130 & 0.37 & 2.08 & 3.20E-01 & 1.47E+01 & 3.58E-01 & 5.66E-03 & 9.08E+00 & 5.07E-03 & 4.20E-04 & 2.25E+00 & 1.33E-02 \\ \hline
        131 & 0.78 & 2.52 & 1.62E-02 & 5.06E+00 & 1.39E-01 & 1.94E-02 & 2.18E+00 & 2.14E-02 & 7.74E-04 & 1.29E+01 & 5.40E-03 \\ \hline
        132 & 0.13 & 0.8 & 2.87E-02 & 1.60E+00 & 3.19E-02 & 2.09E-03 & 2.40E+00 & 1.28E-02 & 1.54E-04 & 1.18E+00 & 1.15E-01 \\ \hline
        133 & 0.15 & 1.02 & 9.04E-03 & 8.39E-01 & 1.13E-01 & 4.64E-03 & 2.44E+00 & 3.83E-02 & 1.12E-03 & 1.95E+00 & 1.02E-01 \\ \hline
        134 & 0.51 & 1.4 & 2.46E-02 & 5.50E+00 & 6.63E-02 & 1.55E-02 & 3.36E+00 & 6.36E-02 & 1.57E-03 & 5.45E+00 & 2.52E-01 \\ \hline
        135 & 0.52 & 1.4 & 2.65E-02 & 2.65E+00 & 4.51E-01 & 2.71E-03 & 1.10E+00 & 8.93E-02 & 2.75E-03 & 4.85E+00 & 1.17E-01 \\ \hline
        136 & 0.32 & 0.5 & 1.67E-03 & 9.10E+00 & 7.05E-03 & 6.39E-03 & 1.29E+01 & 3.04E-04 & 2.88E-03 & 9.64E-01 & 6.60E-01 \\ \hline
        137 & 0 & 0.47 & - & - & - & 8.98E-03 & 7.10E-01 & 4.19E-01 & 2.89E-03 & 4.53E-01 & 2.53E-02 \\ \hline
        138 & 0.12 & 0.26 & 2.62E-02 & 4.43E-01 & 1.09E-01 & 7.61E-02 & 8.68E-01 & 5.84E-01 & 1.15E-03 & 3.05E-01 & 8.62E-02 \\ \hline
        140 & 1 & 0.53 & 2.28E-02 & 1.48E+01 & 4.44E-03 & 3.70E-02 & 2.34E+00 & 4.08E-02 & 5.68E-04 & 2.64E-01 & 4.32E-02 \\ \hline
        141 & 0.341 & 0.55 & 2.37E-02 & 1.24E+00 & 2.30E-01 & 2.74E-03 & 2.70E+00 & 1.23E-01 & 4.51E-04 & 2.90E+00 & 1.29E-01 \\ \hline
        142 & 0.514 & 1.67 & 3.16E-02 & 9.75E+00 & 1.52E+00 & 3.57E-02 & 2.36E+00 & 2.65E-02 & 1.61E-03 & 8.42E-01 & 3.65E-03 \\ \hline
        145 & 0 & 0.89 & - & - & - & 4.92E-03 & 2.02E-01 & 9.22E-02 & 4.77E-04 & 4.31E-01 & 4.75E-02 \\ \hline
        146 & 0.17 & 0.6 & 2.23E-02 & 2.23E+00 & 6.91E-01 & 4.96E-03 & 2.04E+00 & 4.38E-02 & 8.70E-04 & 1.04E+00 & 4.71E-02 \\ \hline
        147 & 0.5 & 2.05 & 1.05E-02 & 1.42E+01 & 1.31E-01 & 4.87E-02 & 1.52E+00 & 4.96E-02 & 1.48E-03 & 3.91E-01 & 1.24E-01 \\ \hline
        148 & 1.02 & 0.78 & 9.50E-03 & 4.29E-01 & 1.42E-01 & 5.95E-03 & 1.43E+00 & 2.89E-01 & 1.01E-02 & 1.14E+01 & 9.50E-01 \\ \hline
        150 & 0.435 & 1.64 & 5.19E-02 & 8.63E+00 & 3.06E+00 & 5.48E-02 & 7.64E+00 & 4.62E-02 & 4.74E-04 & 6.03E+00 & 3.63E-01 \\ \hline
    \end{tabular}
\end{table}
\end{landscape}

\bibliographystyle{spr-mp-sola}
\bibliography{draft1}

\begin{thebibliography}{26}
\ifx\bisbn     \undefined \def\bisbn  #1{ISBN #1}\fi
\ifx\binits    \undefined \def\binits#1{#1}\fi
\ifx\bauthor   \undefined \def\bauthor#1{#1}\fi
\ifx\batitle   \undefined \def\batitle#1{#1}\fi
\ifx\bjtitle   \undefined \def\bjtitle#1{\textit{#1}}\fi
\ifx\bvolume   \undefined \def\bvolume#1{\textbf{#1}}\fi
\ifx\byear     \undefined \def\byear#1{#1}\fi
\ifx\bissue    \undefined \def\bissue#1{#1}\fi
\ifx\bfpage    \undefined \def\bfpage#1{#1}\fi
\ifx\blpage    \undefined \def\blpage #1{#1}\fi
\ifx\burl      \undefined \def\burl#1{#1}\fi
\ifx\href      \undefined \def\href#1#2{#2}\fi
\ifx\betal     \undefined \def\betal{et al.}\fi
\ifx\bctitle   \undefined \def\bctitle#1{#1}\fi
\ifx\beditor   \undefined \def\beditor#1{#1}\fi
\ifx\bbtitle   \undefined \def\bbtitle#1{\textit{#1}}\fi
\ifx\bedition  \undefined \def\bedition#1{#1}\fi
\ifx\bseriesno \undefined \def\bseriesno#1{\textbf{#1}}\fi
\ifx\blocation \undefined \def\blocation#1{#1}\fi
\ifx\bsertitle \undefined \def\bsertitle#1{\textit{#1}}\fi
\ifx\bsnm      \undefined \def\bsnm#1{#1}\fi
\ifx\bsuffix   \undefined \def\bsuffix#1{#1}\fi
\ifx\bparticle \undefined \def\bparticle#1{#1}\fi
\ifx\barticle  \undefined \def\barticle#1{}\fi
\ifx\binstitute  \undefined \def\binstitute#1{#1}\fi
\ifx\bpublisher  \undefined \def\bpublisher#1{#1}\fi
\ifx\doiurl    \undefined \def\doiurl#1{\href{#1}{DOI}}\fi
\makeatletter
\def\safeHref#1#2#3{\in@{http}{#2}\ifin@\href{#2}{#3}\else\href{#1#2}{#3}\fi}
\makeatother
\ifx\adsurl    \undefined
  \def\adsurl#1{\safeHref{https://ui.adsabs.harvard.edu/abs/}{#1}{ADS}}\fi
\ifx\arxivurl  \undefined
  \def\arxivurl#1{\safeHref{http://arxiv.org/abs/}{#1}{arXiv}}\fi
\ifx\botherref \undefined \def\botherref#1{}\fi
\ifx\url       \undefined \def\url#1{#1}\fi
\ifx\bchapter  \undefined \def\bchapter#1{}\fi
\ifx\bbook     \undefined \def\bbook#1{}\fi
\ifx\bcomment  \undefined \def\bcomment#1{#1}\fi
\ifx\oauthor   \undefined \def\oauthor#1{#1}\fi
\ifx\citeauthoryear \undefined\def \citeauthoryear#1{#1}\fi
\def\endbibitem {}
\ifx\bconflocation  \undefined \def\bconflocation#1{#1} \fi

\bibitem[\protect\citeauthoryear{{Bhattacharjee} et~al.}{2023}]{2023Debesh}
\begin{barticle}
\bauthor{\bsnm{{Bhattacharjee}}, \binits{D.}},
\bauthor{\bsnm{{Subramanian}}, \binits{P.}},
\bauthor{\bsnm{{Nieves-Chinchilla}}, \binits{T.}},
\bauthor{\bsnm{{Vourlidas}}, \binits{A.}}:
\byear{2023},
\batitle{{Turbulence and anomalous resistivity inside near-Earth magnetic
  clouds}}.
\bjtitle{\mnras}
\bvolume{518},
\bfpage{1185}.
\doiurl{https://doi.org/10.1093/mnras/stac3186}.
\adsurl{2023MNRAS.518.1185B}.
\end{barticle}
\endbibitem

\bibitem[\protect\citeauthoryear{{Borovsky}}{2012}]{2012borovsky}
\begin{barticle}
\bauthor{\bsnm{{Borovsky}}, \binits{J.E.}}:
\byear{2012},
\batitle{{The velocity and magnetic field fluctuations of the solar wind at 1
  AU: Statistical analysis of Fourier spectra and correlations with plasma
  properties}}.
\bjtitle{Journal of Geophysical Research (Space Physics)}
\bvolume{117},
\bfpage{A05104}.
\doiurl{https://doi.org/10.1029/2011JA017499}.
\adsurl{2012JGRA..117.5104B}.
\end{barticle}
\endbibitem

\bibitem[\protect\citeauthoryear{{Borovsky} and {Funsten}}{2003}]{2003borovsky}
\begin{barticle}
\bauthor{\bsnm{{Borovsky}}, \binits{J.E.}},
\bauthor{\bsnm{{Funsten}}, \binits{H.O.}}:
\byear{2003},
\batitle{{Role of solar wind turbulence in the coupling of the solar wind to
  the Earth's magnetosphere}}.
\bjtitle{Journal of Geophysical Research (Space Physics)}
\bvolume{108},
\bfpage{1246}.
\doiurl{https://doi.org/10.1029/2002JA009601}.
\adsurl{2003JGRA..108.1246B}.
\end{barticle}
\endbibitem

\bibitem[\protect\citeauthoryear{{Bracewell}}{2000}]{2000Bracewell}
\begin{bbook}
\bauthor{\bsnm{{Bracewell}}, \binits{R.N.}}:
\byear{2000},
\bbtitle{{The Fourier transform and its applications}}.
\adsurl{2000fta..book.....B}.
\end{bbook}
\endbibitem

\bibitem[\protect\citeauthoryear{{Bruno} and {Carbone}}{2013}]{2013Bruno}
\begin{barticle}
\bauthor{\bsnm{{Bruno}}, \binits{R.}},
\bauthor{\bsnm{{Carbone}}, \binits{V.}}:
\byear{2013},
\batitle{{The Solar Wind as a Turbulence Laboratory}}.
\bjtitle{Living Reviews in Solar Physics}
\bvolume{10},
\bfpage{2}.
\doiurl{https://doi.org/10.12942/lrsp-2013-2}.
\adsurl{2013LRSP...10....2B}.
\end{barticle}
\endbibitem

\bibitem[\protect\citeauthoryear{{Cid} et~al.}{2002}]{2002cid}
\begin{barticle}
\bauthor{\bsnm{{Cid}}, \binits{C.}},
\bauthor{\bsnm{{Hidalgo}}, \binits{M.A.}},
\bauthor{\bsnm{{Nieves-Chinchilla}}, \binits{T.}},
\bauthor{\bsnm{{Sequeiros}}, \binits{J.}},
\bauthor{\bsnm{{Vi{\~n}as}}, \binits{A.F.}}:
\byear{2002},
\batitle{{Plasma and Magnetic Field Inside Magnetic Clouds: a Global Study}}.
\bjtitle{\solphys}
\bvolume{207},
\bfpage{187}.
\doiurl{https://doi.org/10.1023/A:1015542108356}.
\adsurl{2002SoPh..207..187C}.
\end{barticle}
\endbibitem

\bibitem[\protect\citeauthoryear{{Hu} and {Sonnerup}}{2002}]{2002hu}
\begin{barticle}
\bauthor{\bsnm{{Hu}}, \binits{Q.}},
\bauthor{\bsnm{{Sonnerup}}, \binits{B.U.{\"O}.}}:
\byear{2002},
\batitle{{Reconstruction of magnetic clouds in the solar wind: Orientations and
  configurations}}.
\bjtitle{Journal of Geophysical Research (Space Physics)}
\bvolume{107},
\bfpage{1142}.
\doiurl{https://doi.org/10.1029/2001JA000293}.
\adsurl{2002JGRA..107.1142H}.
\end{barticle}
\endbibitem

\bibitem[\protect\citeauthoryear{{Kilpua}, {Koskinen}, and
  {Pulkkinen}}{2017}]{2017kilpua}
\begin{barticle}
\bauthor{\bsnm{{Kilpua}}, \binits{E.}},
\bauthor{\bsnm{{Koskinen}}, \binits{H.E.J.}},
\bauthor{\bsnm{{Pulkkinen}}, \binits{T.I.}}:
\byear{2017},
\batitle{{Coronal mass ejections and their sheath regions in interplanetary
  space}}.
\bjtitle{Living Reviews in Solar Physics}
\bvolume{14},
\bfpage{5}.
\doiurl{https://doi.org/10.1007/s41116-017-0009-6}.
\adsurl{2017LRSP...14....5K}.
\end{barticle}
\endbibitem

\bibitem[\protect\citeauthoryear{Kilpua et~al.}{2013}]{2013kilpua}
\begin{bchapter}
\bauthor{\bsnm{Kilpua}, \binits{E.}},
\bauthor{\bsnm{Hietala}, \binits{H.}},
\bauthor{\bsnm{Koskinen}, \binits{H.}},
\bauthor{\bsnm{Fontaine}, \binits{D.}},
\bauthor{\bsnm{Turc}, \binits{L.}}:
\byear{2013},
\bctitle{Magnetic field and dynamic pressure ULF fluctuations in
  coronal-mass-ejection-driven sheath regions}.
In: \bbtitle{Annales Geophysicae}
\bseriesno{31},
\bfpage{1559}.
\bcomment{Copernicus Publications G{\"o}ttingen, Germany}.
\end{bchapter}
\endbibitem

\bibitem[\protect\citeauthoryear{Kilpua et~al.}{2019}]{2019kilpuasolar}
\begin{barticle}
\bauthor{\bsnm{Kilpua}, \binits{E.}},
\bauthor{\bsnm{Fontaine}, \binits{D.}},
\bauthor{\bsnm{Moissard}, \binits{C.}},
\bauthor{\bsnm{Ala-Lahti}, \binits{M.}},
\bauthor{\bsnm{Palmerio}, \binits{E.}},
\bauthor{\bsnm{Yordanova}, \binits{E.}},
\bauthor{\bsnm{Good}, \binits{S.}},
\bauthor{\bsnm{Kalliokoski}, \binits{M.}},
\bauthor{\bsnm{Lumme}, \binits{E.}},
\bauthor{\bsnm{Osmane}, \binits{A.}}, \betal:
\byear{2019},
\batitle{Solar wind properties and geospace impact of coronal mass
  ejection-driven sheath regions: Variation and driver dependence}.
\bjtitle{Space Weather}
\bvolume{17},
\bfpage{1257}.
\end{barticle}
\endbibitem

\bibitem[\protect\citeauthoryear{{Klein} and {Burlaga}}{1982}]{1982Klein}
\begin{barticle}
\bauthor{\bsnm{{Klein}}, \binits{L.W.}},
\bauthor{\bsnm{{Burlaga}}, \binits{L.F.}}:
\byear{1982},
\batitle{{Interplanetary magnetic clouds at 1 AU}}.
\bjtitle{\jgr}
\bvolume{87},
\bfpage{613}.
\doiurl{https://doi.org/10.1029/JA087iA02p00613}.
\adsurl{1982JGR....87..613K}.
\end{barticle}
\endbibitem

\bibitem[\protect\citeauthoryear{{Larrodera} and
  {Temmer}}{2024}]{2024Larrodera}
\begin{barticle}
\bauthor{\bsnm{{Larrodera}}, \binits{C.}},
\bauthor{\bsnm{{Temmer}}, \binits{M.}}:
\byear{2024},
\batitle{{Evolution of coronal mass ejections with and without sheaths from the
  inner to the outer heliosphere: Statistical investigation for 1975 to 2022}}.
\bjtitle{\aap}
\bvolume{685},
\bfpage{A89}.
\doiurl{https://doi.org/10.1051/0004-6361/202348641}.
\adsurl{2024A&A...685A..89L}.
\end{barticle}
\endbibitem

\bibitem[\protect\citeauthoryear{{Lazarian} and
  {Vishniac}}{1999}]{1999Lazarian}
\begin{barticle}
\bauthor{\bsnm{{Lazarian}}, \binits{A.}},
\bauthor{\bsnm{{Vishniac}}, \binits{E.T.}}:
\byear{1999},
\batitle{{Reconnection in a Weakly Stochastic Field}}.
\bjtitle{\apj}
\bvolume{517},
\bfpage{700}.
\doiurl{https://doi.org/10.1086/307233}.
\adsurl{1999ApJ...517..700L}.
\end{barticle}
\endbibitem

\bibitem[\protect\citeauthoryear{{Liu} et~al.}{2006}]{2006Liu}
\begin{barticle}
\bauthor{\bsnm{{Liu}}, \binits{Y.}},
\bauthor{\bsnm{{Richardson}}, \binits{J.D.}},
\bauthor{\bsnm{{Belcher}}, \binits{J.W.}},
\bauthor{\bsnm{{Kasper}}, \binits{J.C.}},
\bauthor{\bsnm{{Elliott}}, \binits{H.A.}}:
\byear{2006},
\batitle{{Thermodynamic structure of collision-dominated expanding plasma:
  Heating of interplanetary coronal mass ejections}}.
\bjtitle{Journal of Geophysical Research (Space Physics)}
\bvolume{111},
\bfpage{A01102}.
\doiurl{https://doi.org/10.1029/2005JA011329}.
\adsurl{2006JGRA..111.1102L}.
\end{barticle}
\endbibitem

\bibitem[\protect\citeauthoryear{{Manoharan} et~al.}{2000}]{2000Manoharan}
\begin{barticle}
\bauthor{\bsnm{{Manoharan}}, \binits{P.K.}},
\bauthor{\bsnm{{Kojima}}, \binits{M.}},
\bauthor{\bsnm{{Gopalswamy}}, \binits{N.}},
\bauthor{\bsnm{{Kondo}}, \binits{T.}},
\bauthor{\bsnm{{Smith}}, \binits{Z.}}:
\byear{2000},
\batitle{{Radial Evolution and Turbulence Characteristics of a Coronal Mass
  Ejection}}.
\bjtitle{\apj}
\bvolume{530},
\bfpage{1061}.
\doiurl{https://doi.org/10.1086/308378}.
\adsurl{2000ApJ...530.1061M}.
\end{barticle}
\endbibitem

\bibitem[\protect\citeauthoryear{{Mitsakou} and {Moussas}}{2014}]{2014Mitsakou}
\begin{barticle}
\bauthor{\bsnm{{Mitsakou}}, \binits{E.}},
\bauthor{\bsnm{{Moussas}}, \binits{X.}}:
\byear{2014},
\batitle{{Statistical Study of ICMEs and Their Sheaths During Solar Cycle 23
  (1996 - 2008)}}.
\bjtitle{\solphys}
\bvolume{289},
\bfpage{3137}.
\doiurl{https://doi.org/10.1007/s11207-014-0505-y}.
\adsurl{2014SoPh..289.3137M}.
\end{barticle}
\endbibitem

\bibitem[\protect\citeauthoryear{{M{\"o}stl} et~al.}{2009}]{2009Mostl}
\begin{barticle}
\bauthor{\bsnm{{M{\"o}stl}}, \binits{C.}},
\bauthor{\bsnm{{Farrugia}}, \binits{C.J.}},
\bauthor{\bsnm{{Biernat}}, \binits{H.K.}},
\bauthor{\bsnm{{Leitner}}, \binits{M.}},
\bauthor{\bsnm{{Kilpua}}, \binits{E.K.J.}},
\bauthor{\bsnm{{Galvin}}, \binits{A.B.}},
\bauthor{\bsnm{{Luhmann}}, \binits{J.G.}}:
\byear{2009},
\batitle{{Optimized Grad - Shafranov Reconstruction of a Magnetic Cloud Using
  STEREO- Wind Observations}}.
\bjtitle{\solphys}
\bvolume{256},
\bfpage{427}.
\doiurl{https://doi.org/10.1007/s11207-009-9360-7}.
\adsurl{2009SoPh..256..427M}.
\end{barticle}
\endbibitem

\bibitem[\protect\citeauthoryear{{Narock} et~al.}{2024}]{2024Narock}
\begin{barticle}
\bauthor{\bsnm{{Narock}}, \binits{T.}},
\bauthor{\bsnm{{Pal}}, \binits{S.}},
\bauthor{\bsnm{{Arsham}}, \binits{A.}},
\bauthor{\bsnm{{Narock}}, \binits{A.}},
\bauthor{\bsnm{{Nieves-Chinchilla}}, \binits{T.}}:
\byear{2024},
\batitle{{Classifying Different Types of Solar-Wind Plasma with Uncertainty
  Estimations Using Machine Learning}}.
\bjtitle{\solphys}
\bvolume{299},
\bfpage{131}.
\doiurl{https://doi.org/10.1007/s11207-024-02379-8}.
\adsurl{2024SoPh..299..131N}.
\end{barticle}
\endbibitem

\bibitem[\protect\citeauthoryear{{Nieves-Chinchilla} et~al.}{2018}]{2018NCSo}
\begin{barticle}
\bauthor{\bsnm{{Nieves-Chinchilla}}, \binits{T.}},
\bauthor{\bsnm{{Vourlidas}}, \binits{A.}},
\bauthor{\bsnm{{Raymond}}, \binits{J.C.}},
\bauthor{\bsnm{{Linton}}, \binits{M.G.}},
\bauthor{\bsnm{{Al-haddad}}, \binits{N.}},
\bauthor{\bsnm{{Savani}}, \binits{N.P.}},
\bauthor{\bsnm{{Szabo}}, \binits{A.}},
\bauthor{\bsnm{{Hidalgo}}, \binits{M.A.}}:
\byear{2018},
\batitle{{Understanding the Internal Magnetic Field Configurations of ICMEs
  Using More than 20 Years of Wind Observations}}.
\bjtitle{\solphys}
\bvolume{293},
\bfpage{25}.
\doiurl{https://doi.org/10.1007/s11207-018-1247-z}.
\adsurl{2018SoPh..293...25N}.
\end{barticle}
\endbibitem

\bibitem[\protect\citeauthoryear{{Nieves-Chinchilla} et~al.}{2019}]{2019NCSoPh}
\begin{barticle}
\bauthor{\bsnm{{Nieves-Chinchilla}}, \binits{T.}},
\bauthor{\bsnm{{Jian}}, \binits{L.K.}},
\bauthor{\bsnm{{Balmaceda}}, \binits{L.}},
\bauthor{\bsnm{{Vourlidas}}, \binits{A.}},
\bauthor{\bsnm{{dos Santos}}, \binits{L.F.G.}},
\bauthor{\bsnm{{Szabo}}, \binits{A.}}:
\byear{2019},
\batitle{{Unraveling the Internal Magnetic Field Structure of the
  Earth-directed Interplanetary Coronal Mass Ejections During 1995 - 2015}}.
\bjtitle{\solphys}
\bvolume{294},
\bfpage{89}.
\doiurl{https://doi.org/10.1007/s11207-019-1477-8}.
\adsurl{2019SoPh..294...89N}.
\end{barticle}
\endbibitem

\bibitem[\protect\citeauthoryear{Richardson and Cane}{2024}]{richcane_2024}
\begin{botherref}
\oauthor{\bsnm{Richardson}, \binits{I.}},
\oauthor{\bsnm{Cane}, \binits{H.}}:
2024,
{Near-Earth Interplanetary Coronal Mass Ejections Since January 1996}.
\doiurl{https://doi.org/10.7910/DVN/C2MHTH}.
\url{https://doi.org/10.7910/DVN/C2MHTH}.
\end{botherref}
\endbibitem

\bibitem[\protect\citeauthoryear{{Sachdeva} et~al.}{2017}]{2017Nishtha}
\begin{barticle}
\bauthor{\bsnm{{Sachdeva}}, \binits{N.}},
\bauthor{\bsnm{{Subramanian}}, \binits{P.}},
\bauthor{\bsnm{{Vourlidas}}, \binits{A.}},
\bauthor{\bsnm{{Bothmer}}, \binits{V.}}:
\byear{2017},
\batitle{{CME Dynamics Using STEREO and LASCO Observations: The Relative
  Importance of Lorentz Forces and Solar Wind Drag}}.
\bjtitle{\solphys}
\bvolume{292},
\bfpage{118}.
\doiurl{https://doi.org/10.1007/s11207-017-1137-9}.
\adsurl{2017SoPh..292..118S}.
\end{barticle}
\endbibitem

\bibitem[\protect\citeauthoryear{{Salman} et~al.}{2020}]{2020salman}
\begin{barticle}
\bauthor{\bsnm{{Salman}}, \binits{T.M.}},
\bauthor{\bsnm{{Lugaz}}, \binits{N.}},
\bauthor{\bsnm{{Farrugia}}, \binits{C.J.}},
\bauthor{\bsnm{{Winslow}}, \binits{R.M.}},
\bauthor{\bsnm{{Jian}}, \binits{L.K.}},
\bauthor{\bsnm{{Galvin}}, \binits{A.B.}}:
\byear{2020},
\batitle{{Properties of the Sheath Regions of Coronal Mass Ejections with or
  without Shocks from STEREO in situ Observations near 1 au}}.
\bjtitle{\apj}
\bvolume{904},
\bfpage{177}.
\doiurl{https://doi.org/10.3847/1538-4357/abbdf5}.
\adsurl{2020ApJ...904..177S}.
\end{barticle}
\endbibitem

\bibitem[\protect\citeauthoryear{{Sorriso-Valvo} et~al.}{2021}]{2021Sorriso}
\begin{barticle}
\bauthor{\bsnm{{Sorriso-Valvo}}, \binits{L.}},
\bauthor{\bsnm{{Yordanova}}, \binits{E.}},
\bauthor{\bsnm{{Dimmock}}, \binits{A.P.}},
\bauthor{\bsnm{{Telloni}}, \binits{D.}}:
\byear{2021},
\batitle{{Turbulent Cascade and Energy Transfer Rate in a Solar Coronal Mass
  Ejection}}.
\bjtitle{\apjl}
\bvolume{919},
\bfpage{L30}.
\doiurl{https://doi.org/10.3847/2041-8213/ac26c5}.
\adsurl{2021ApJ...919L..30S}.
\end{barticle}
\endbibitem

\bibitem[\protect\citeauthoryear{{Temmer} and {Bothmer}}{2022}]{2022Temmer}
\begin{barticle}
\bauthor{\bsnm{{Temmer}}, \binits{M.}},
\bauthor{\bsnm{{Bothmer}}, \binits{V.}}:
\byear{2022},
\batitle{{Characteristics and evolution of sheath and leading edge structures
  of interplanetary coronal mass ejections in the inner heliosphere based on
  Helios and Parker Solar Probe observations}}.
\bjtitle{\aap}
\bvolume{665},
\bfpage{A70}.
\doiurl{https://doi.org/10.1051/0004-6361/202243291}.
\adsurl{2022A&A...665A..70T}.
\end{barticle}
\endbibitem

\bibitem[\protect\citeauthoryear{Wand}{1997}]{1997Wand01021997}
\begin{barticle}
\bauthor{\bsnm{Wand}, \binits{M.P.}}:
\byear{1997},
\batitle{Data-Based Choice of Histogram Bin Width}.
\bjtitle{The American Statistician}
\bvolume{51},
\bfpage{59}.
\doiurl{https://doi.org/10.1080/00031305.1997.10473591}.
\burl{https://www.tandfonline.com/doi/abs/10.1080/00031305.1997.10473591}.
\end{barticle}
\endbibitem

\end{thebibliography}

\end{document}